\documentclass[aps,prx,reprint,superscriptaddress]{revtex4-2}
\usepackage{graphicx}
\usepackage{natbib}
\usepackage{siunitx}
\usepackage{soul,xcolor}
\usepackage{nicefrac}
\usepackage{amsmath}
\usepackage{cancel}

\begin{document}

\title{Magnetic frustration in a metallic fcc lattice}

\author{Oliver Stockert}
\email{oliver.stockert@cpfs.mpg.de}
\affiliation{Max-Planck-Institut f\"ur Chemische Physik fester Stoffe, 01187 Dresden, Germany}
\author{Jens-Uwe Hoffmann}
\affiliation{Helmholtz-Zentrum Berlin f\"ur Materialien und Energie, 14109 Berlin, Germany}
\author{Martin M\"uhlbauer}
\author{Anatoliy Senyshyn}
\affiliation{Heinz Maier-Leibnitz Zentrum, 85747 Garching, Germany}
\author{Michael~M.~Koza}
\affiliation{Institut Laue-Langevin, 38042 Grenoble, France}
\author{Alexander A. Tsirlin}
\author{F. Maximilian Wolf}
\author{Sebastian Bachus}
\author{Philipp Gegenwart}
\affiliation{Experimental Physics VI, Center for Electronic Correlations and Magnetism, Institute of Physics, University of Augsburg, 86135 Augsburg, Germany}
\author{Roman Movshovich}
\affiliation{MPA-CMMS, Los Alamos National Laboratory, Los Alamos, New Mexico 87545, USA}
\author{Svilen Bobev}
\affiliation{Department of Chemistry and Biochemistry, University of Delaware, Newark, DE 19716, USA}
\author{Veronika Fritsch}
\email{veronika.fritsch@physik.uni-augsburg.de}
\affiliation{Experimental Physics VI, Center for Electronic Correlations and Magnetism, Institute of Physics, University of Augsburg, 86135 Augsburg, Germany}

\date{\today}

\begin{abstract}
Magnetic frustration in metals is scarce and hard to pinpoint, but exciting due to the possibility of the emergence of fascinating novel phases. The cubic intermetallic compound HoInCu$_4$ with all holmium atoms on an fcc lattice, exhibits partial magnetic frustration, yielding a ground state where half of the Ho moments remain without long-range order, as evidenced by our neutron scattering experiments. The substitution of In with Cd results in HoCdCu$_4$ in a full breakdown of magnetic
frustration. Consequently we found a fully ordered magnetic structure in our neutron diffraction experiments. These findings are in agreement with the local energy scales and crystal electric field excitations, which we determined from specific heat and inelastic neutron scattering data.
The electronic density of states for the itinerant bands acts as tuning parameter for the ratio between nearest-neighbor and next-nearest-neighbor interactions and thus for magnetic frustration.
\end{abstract}

\pacs{}

\maketitle

\section{Introduction}
Effects of magnetic frustration on the magnetic properties of compounds are in the focus of current condensed matter research. As a consequence of frustration exotic ground states can occur, e.g. spin-glass, spin-liquid or spin-ice states \cite{Balents2010,Lacroix2011}.
The prototypes of geometrically frustrated systems in two dimensions are the triangular lattice and the kagome lattice and their three-dimensional counterparts the face-centered cubic (fcc) lattice and the pyrochlore lattice \cite{Ramirez1994Strongly}. While the kagome lattice with the lowest coordination number of the aforementioned is most prone to magnetic frustration, the fcc lattice with its three times higher coordination number exhibits magnetic frustration less often.
An example for a geometrical frustrated fcc lattice is K$_2$IrCl$_6$ \cite{Cooke1959Exchange,Ramirez1994Strongly,Khan2019Cubic}. Another  example is the Kitaev model for an fcc lattice \cite{Kimchi2014Kitaev}, featuring for example an unusually large magnon gap caused by quantum order-by-disorder in La$_2B$IrO$_6$ with $B =$ Mg, Zn \cite{Aczel2016Highly}. In both cases a strong spin-orbit coupling (SOC) is an essential ingredient for the emerging unusual states.

A large SOC is as well found in heavy rare-earth compounds, e.g. in Ho alloys: The half-Heusler compound HoPdSb is characterized by its thermoelectric properties, however, despite its very low conductivity and its rather low N\'{e}el temperature $T_\mathrm{N} = 2.2\,\si{K}$ there are no clear signs of magnetic frustration \cite{Gofryk2005Magnetic,Mukhopadhyay2018Multi}.
In the Shastry-Sutherland lattice of HoB$_4$ the ground state is determined by RKKY interactions in zero field \cite{Okuyama2007Magnetic}, in a finite field, however, geometrical frustration of quadrupolar interactions emerges \cite{Kim2009Anisotropic}. HoB$_{12}$ again with an fcc arrangement of the magnetic Ho atoms, displays a rich magnetic $B - T$ phase diagram and orders in an incommensurate antiferromagnetic structure with a propagation vector $(1/2-\delta~1/2-\delta~1/2-\delta)$ \cite{Kohout2004,Siemensmeyer2007}. In contrast, the Laves phase HoNi$_2$, where Ni is non-magnetic, orders ferromagnetically at $T_C \approx 12\,\si{K}$ \cite{Gomes2003Low}. HoCu$_5$, being isostructural to the here investigated HoInCu$_4$, is reported to exhibit ferromagnetic order close to a magnetic instability \cite{Buschow1970Magnetic}.

In contrast to insulating materials magnetic frustration in metallic systems is quite scarce due to the more long-ranged interactions mediated by the conduction electrons and the possible delocalization of magnetic moments due to the Kondo effect. One of the first intermetallic systems identified as a partially frustrated metal is GdInCu$_4$, where the Gd ions form an fcc lattice: the magnetic structure of GdInCu$_4$ below $T_{\rm N} \approx 7$\,K was determined as consisting of antiferromagnetic planes perpendicular to $\left<100\right>$, which are stacked antiferromagnetically to each other and separated by frustrated planes \cite{Nakamura1993Anomalous,Nakamura1999partially}. In the isostructural compounds $R$InCu$_4$ with $R$ being Dy, Ho or Er large frustration parameters $f:=\frac{-\theta_\mathrm{CW}}{T_\mathrm{N}}$ with values $f > 10$ were found \cite{Fritsch2005Spin}. The substitution of the non-magnetic In with non-magnetic Cd yields a dramatic decrease of the frustration parameter indicating a breakdown of the magnetic frustration. This vanishing of frustration is accompanied by a significant qualitative and quantitative change of the resistivity from a concavely shaped curve of a bad metal resistivity into a linear resistivity with a conductivity one order of magnitude improved \cite{Fritsch2006Correlation}.

Here we present detailed investigations on the magnetic structure and the thermodynamic properties of HoInCu$_4$, which is partially frustrated and orders antiferromagnetically only below $T_{\rm N} = 0.76$\,K. We compare our results to HoCdCu$_4$, which exhibits a fully ordered magnetic structure below $T_{\rm N} \approx 7$\,K. A theoretical description is based on
the antiferromagnetic $J_1-J_2$ Heisenberg model on an fcc lattice which has been extensively studied \cite{Lines1964Green,Lacroix2011,Balents2010,Sun2018J1}. Depending on the ratio of next-nearest-neighbor to nearest-neighbor interactions $J_2/J_1$ several types of ferro- and antiferromagnetic phases are predicted. Our data indicate HoInCu$_4$ is located very close to a magnetic instability, making it a promising candidate on the quest for new exotic ground states. For $J_2/J_1 \geq 0.5$ the magnetic structure predicted is the same as we found for HoCdCu$_4$.

\section{Experimental Details and Crystallographic Structure}
HoInCu$_4$ and HoCdCu$_4$ single crystals
were grown by the same flux-growth method as reported previously \cite{Fritsch2005Spin,Fritsch2006Correlation}. Powder samples were obtained by grinding single crystals. Single crystal x-ray diffraction data were recorded on a Bruker SMART APEX CCD diffractometer equipped with a state-of-the-art low-temperature apparatus ($120\,\si{K}$) and using graphite monochromatized Mo K$_\alpha$ radiation. Although the existence of the cubic phase HoInCu$_4$ already was reported previously \cite{Nakamura1995Transport,Fritsch2005Spin}, here we provide the first structural analysis and deposited it in the structure database \cite{CCDCFIZ}, See Supplemental Material \cite{Supplemental} for the experimental setup and the full results.  The Rietveld-refinement of a powder x-ray diffraction pattern of HoInCu$_4$ is shown in the Supplemental Material \cite{Supplemental}  as well. Our single-crystal and powder diffraction data are in excellent agreement and reveal a fully ordered structure. Neither between the Ho and the In sites nor between the Cu and the In sites were found any signs of disorder. Note that In and Cu have sufficiently different x-ray scattering factors, which make them clearly distinguishable. The absence of a detectable site disorder in HoInCu$_4$ justifies to view the system as two networks of corner-sharing tetrahedra of Ho-ions, where the tetrahedra in one network are filled with In ions, in the other with tetrahedra of Cu-ions \cite{Fritsch2005Spin}.

Heat capacity measurements down to $2\,\si{K}$ were conducted in a Physical Property Measurement System (PPMS, Quantum Design), while for measurements down to $70\,\si{mK}$ a quasiadiabatic thermal relaxation method in a dilution refrigerator was utilized.

Powder and single crystal neutron diffraction on HoInCu$_4$ was carried out on the flat-cone diffractometer E2 at BER-II at HZB/Berlin using dilution and $^3$He-cryostats. Measurements were performed at temperatures between $T = 65$\,mK and $50$\,K with a neutron wavelength $\lambda = 2.39$\,{\AA}. For the powder diffraction experiment about 8\,g of HoInCu$_4$ powder was filled in a sealed copper sample container together with some deuterated methanol-ethanol mixture to improve the thermal coupling of the powder at low temperatures (below $\approx 1$\,K). The same 8\,g of HoInCu$_4$ powder served for the inelastic neutron scattering measurements on the time-of-flight spectrometer IN6 at the high-flux reactor of the ILL/Grenoble. Here the incident neutron energy was fixed to $E_i = 4.77$\,meV. The powder was filled in a disk-shaped container with a thickness of $<1$\,mm to minimize neutron absorption. Data were recorded between $1.6$ and $100$\,K in a $^4$He Orange cryostat to mainly study the crystalline electric field excitations. A HoInCu$_4$ single crystal ($m \approx 630$\,mg) was used for the single crystal diffraction and mounted with the $[001]$ axis vertical resulting in a $(h~k~0)$ horizontal scattering plane.

Neutron diffraction on HoCdCu$_4$ powder was performed using the the diffractometer SPODI at MLZ/Garching with a neutron wavelength $\lambda = 1.5483$\,{\AA}. The HoCdCu$_4$ powder was filled in the space between two concentric aluminium cylinders with 0.5 \,mm distance to ensure a reduced neutron absorption. The sample was mounted on the cold head of a closed-cycle refrigerator. Powder pattern were recorded up to $2\Theta \approx 150$\,degs. at 300, 20 , and 4\,K, i.e. above and below the N\'eel temperature $T_{\rm N} = 7.0$\,K. For all neutron scattering data the error bars denote the statistical error
($\pm 1 \sigma$ interval).

\section{Characteristic energy scales}
Thermodynamic measurements including heat capacity and magnetic susceptibility as well as inelastic neutron scattering experiments have been carried out to study the characteristic energy scales of the magnetic Ho atoms in HoInCu$_4$. In particular, we focused on the determination of the crystalline-electric-field (CEF) level scheme of the Ho 4f moments and the hyperfine splitting of the nuclear Ho moments. Knowing these local energy scales helps in the understanding of the magnetic order  of HoInCu$_4$ below $T_{\rm N} = 0.76$\,K.

\begin{figure}
\centering
\includegraphics[width=0.95\columnwidth]{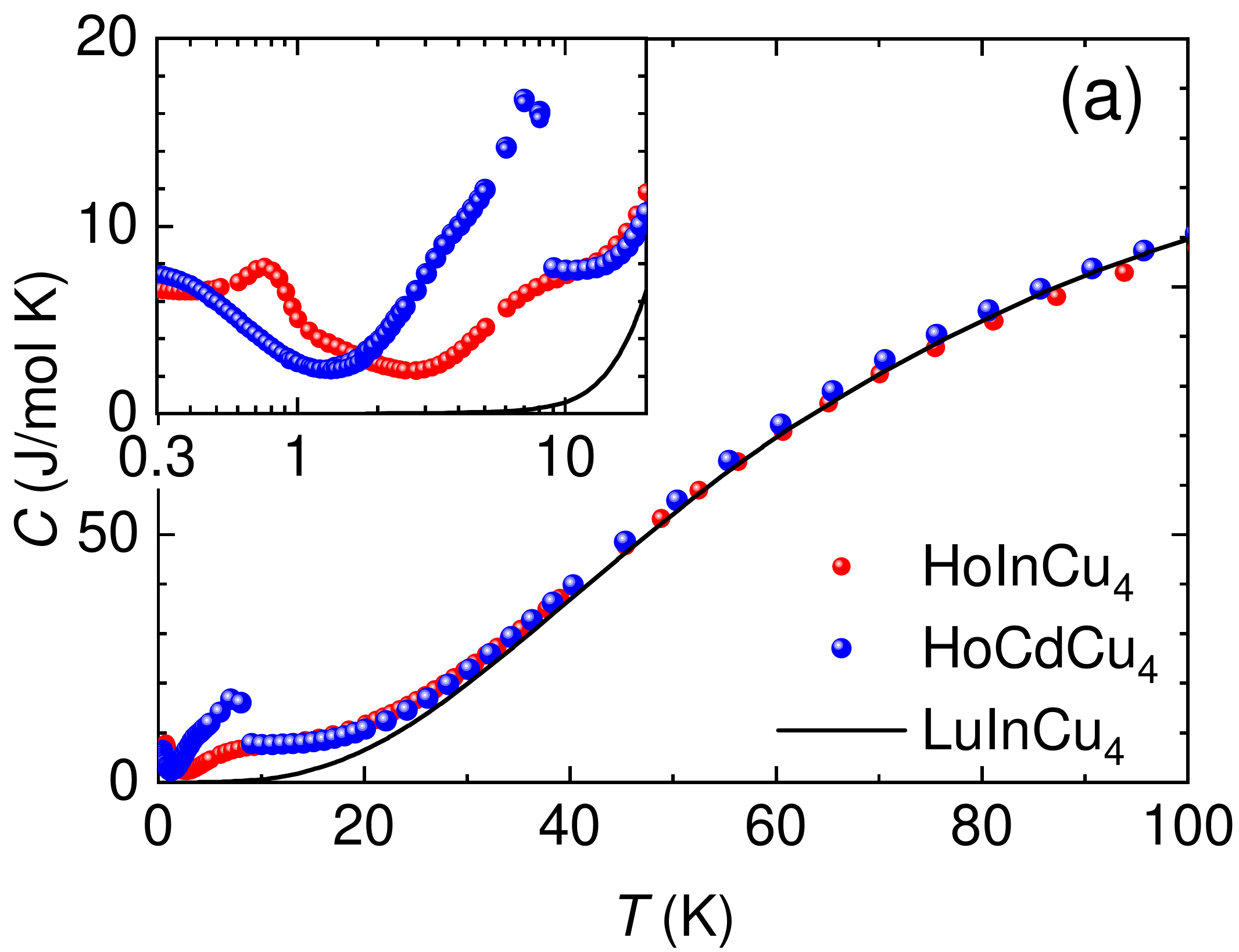}%
\vskip3ex
\includegraphics[width=0.95\columnwidth]{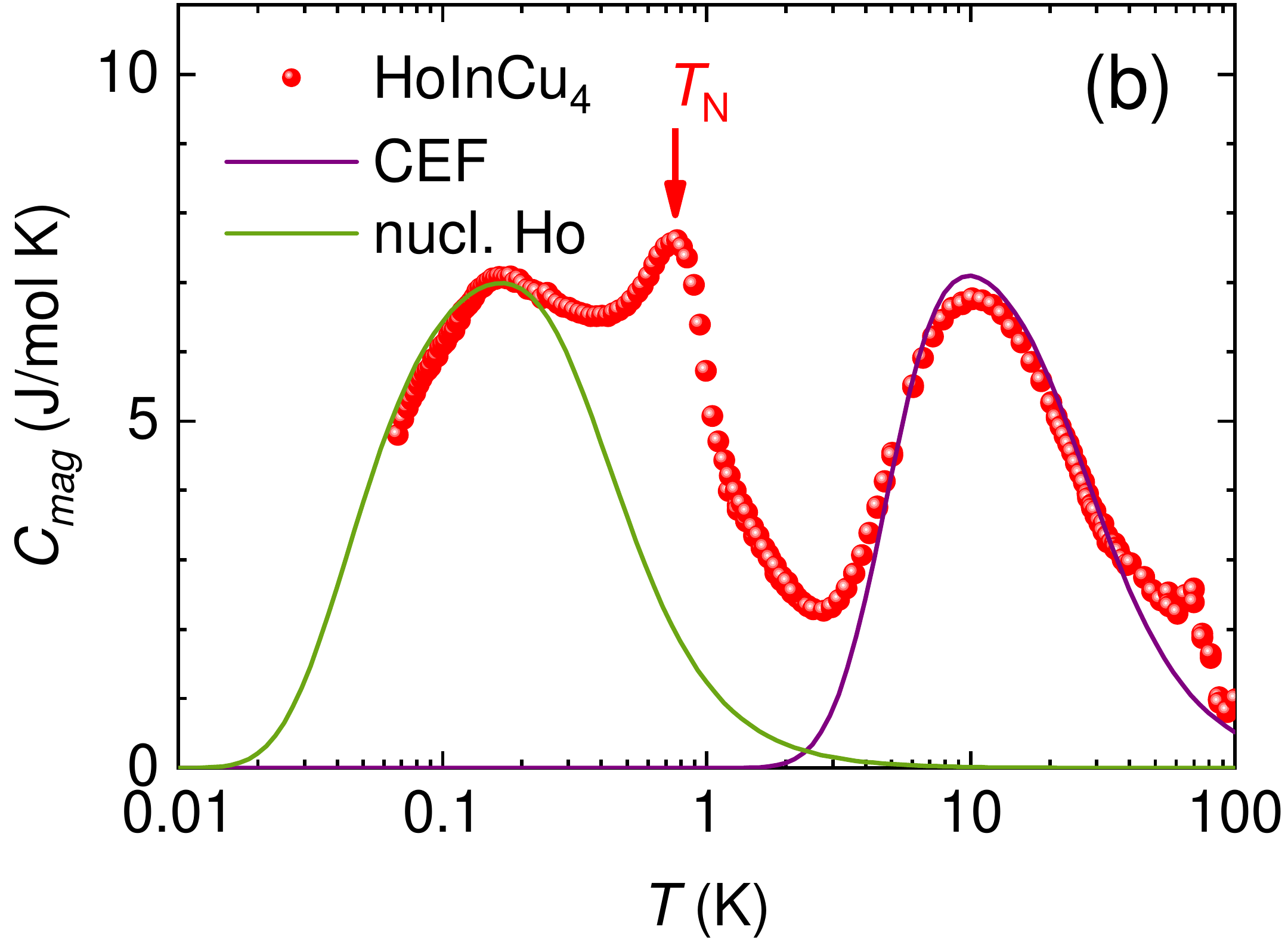}
\vskip3ex
\includegraphics[width=0.95\columnwidth]{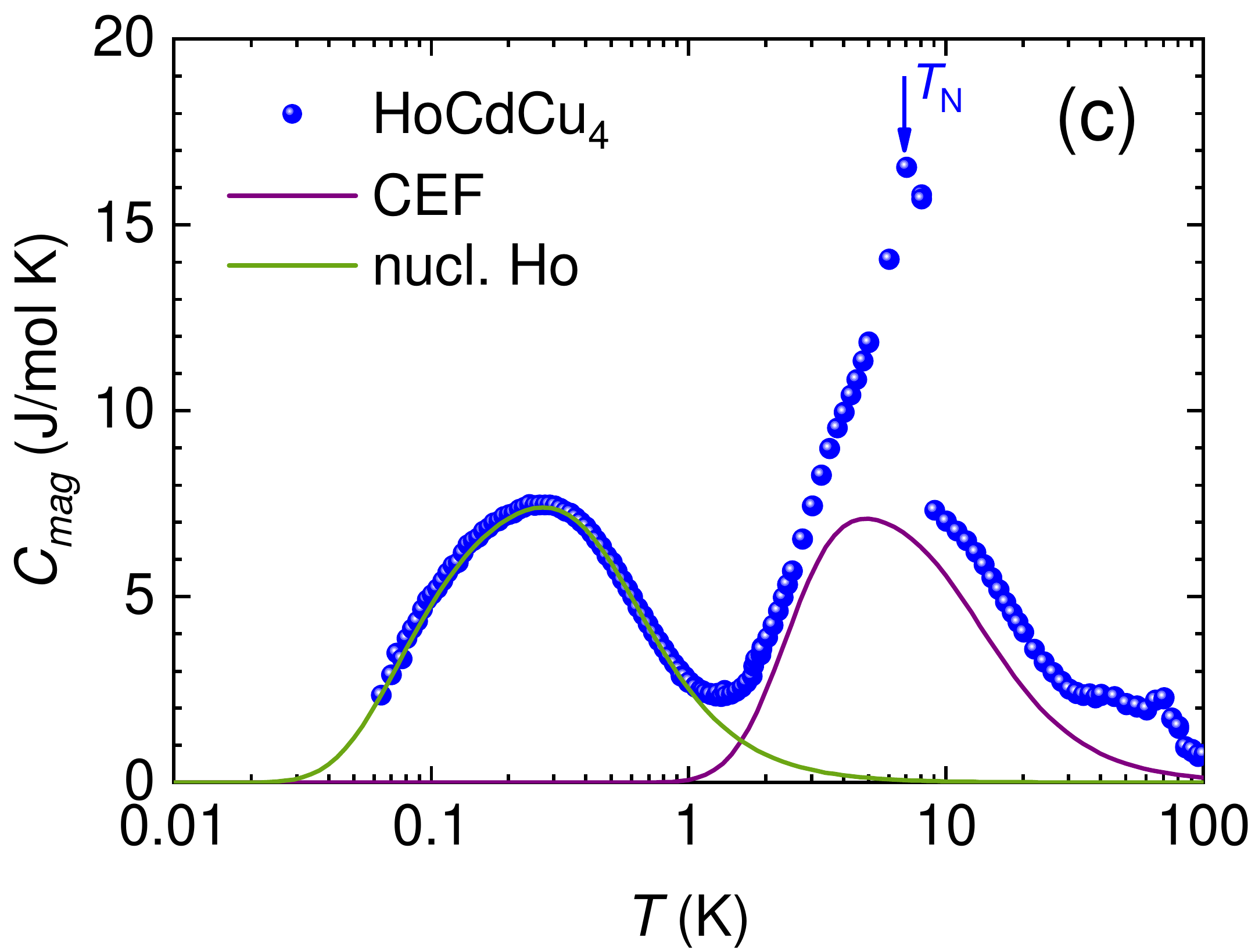}
\caption{(a) Temperature dependence of the total heat capacity of HoInCu$_4$, HoCdCu$_4$ and LuInCu$_4$ up to $100$\,K. Note that the heat capacity of LuInCu$_4$ exceeds the one of HoInCu$_4$ above $\approx 60$\,K. The inset enlarges the low-temperature regime below $T = 20$\,K. (b+c) Magnetic heat capacity of HoInCu$_4$ (b) and HoCdCu$_4$ (c) after subtraction of the phonon contribution. Solid lines indicate the contribution of the CEF excitations to the heat capacity (purple line) and the heat capacity due to excitations of the nuclear magnetic moments (green line). Data partially from refs.~\cite{Fritsch2005Spin,Fritsch2006Correlation}. }\label{FullHC}
\end{figure}

The temperature dependence of the heat capacity for HoInCu$_4$ and HoCdCu$_4$ is plotted in Fig.\,\ref{FullHC}(a).
When analyzing the heat capacity data in order to extract the magnetic contribution of the Ho 4f electrons, a few important issues have to be considered: the phonon contribution to the total heat capacity and the presence of a term due to the nuclear moment of holmium and its effect on the heat capacity due to the strong hyperfine coupling. Holmium is located close to a full occupation of the 4f shell. Therefore, LuInCu$_4$ was used as a nonmagnetic parent compound to determine the phonon contribution to the heat capacity. However, since the atomic masses of holmium and lutetium differ by roughly $10$\,amu, a direct subtraction of the measured heat capacity $C$ of LuInCu$_4$ from the HoInCu$_4$ data is impossible. As seen in Fig.\,\ref{FullHC}(a), $C_{\rm LuInCu_4}$ even exceeds $C_{\rm HoInCu_4}$ above $\approx 60$\,K. For a better estimation of the phonon contribution to the heat capacity the specific heat of LuInCu$_4$ has been rescaled by the mass differences of the formula units. Within the Debye approximation which is valid at low temperatures far below the Debye temperature $\Theta_D$, the specific heat varies as $C \propto (T/\Theta_D)^3$ with $\Theta_D$ being inverse proportional to the square-root of the mass of one formula unit, here LuInCu$_4$ or HoInCu$_4$. Rescaling in this way the specific heat of LuInCu$_4$ to get an estimation of the phonon contribution in the total specific heat of HoInCu$_4$, yields a scale factor of $0.972$ by which $C_{\rm LuInCu_4}$ has to be multiplied. In this simple approach the magnetic specific heat of HoInCu$_4$, $C_{\rm mag}$, is then obtained by $C_{\rm mag} = C_{\rm HoInCu_4} - 0.972\cdot C_{\rm LuInCu_4}$. As seen in Fig.\,\ref{FullHC}(b) three peaks can be clearly distinguished in the temperature dependence of $C_{\rm mag}$: (i) a Schottky-like anomaly at around $10$\,K which can be nicely described by the CEF parameters obtained from INS (purple solid line), (ii) a maximum below $1$\,K indicating the onset of magnetic order of the holmium 4f moments, and (iii) a peak just below $200$\,mK originating from a nuclear Schottky anomaly due to the splitting of the nuclear spins of holmium in the field produced by the 4f moments. Since only one holmium isotope exists, $^{165}$Ho with a nuclear spin $I = 7/2$, the eightfold degenerate nuclear ground state splits in the presence of static 4f moments. The Hamiltonian for the hyperfine interactions can be written as $H_{hfi} = aI_z + p\left[I_z^2 - 1/3 I (I+1)\right]$ \cite{Lounasmaa1962Specific}. The first term is the hyperfine coupling of the nuclear moment with the 4f moment, while the second term denotes the coupling of the nuclear quadrupole moment with the electric field gradient. The energy eigenvalues of $H_{hfi}$ are given by $E_m = am + p{m^2 - 1/3 I (I+1)}$ with $-I \le m\le I$. When neglecting the quadrupole term, the energy eigenvalues $E_m$ are equally spaced. The nuclear heat capacity is then calculated from the energy splittings $E_m$. To describe the experimental specific heat at low $T$, $a$ and $p$ were varied and it was looked for the best match of the calculated nuclear heat capacity and the experimental data. As a result, $a = 0.18$\,K($\simeq 1.55\cdot10^{-2}$\,meV) and $p = 4\cdot10^{-3}$\,K ($\simeq 3.45\cdot10^{-4}$\,meV) were obtained as best parameters. It should be noted that the literature values for $a$ and $p$ cannot describe our data satisfactorily \cite{Lounasmaa1962Specific}. In contrast to HoInCu$_4$ where $C_{\rm mag}$ peaks well below $200$\,mK, all other Ho compounds reported in literature, show a larger energy splitting of the nuclear moments, i.e. a nuclear Schottky anomaly at much higher temperature \cite{Lounasmaa1962Specific,Ehlers2009,Chatterji2013nuclear,Chatterji2013direct,Kumar2016}. The reduced temperature scale of the Schottky anomaly in HoInCu$_4$ is attributed to the reduced internal magnetic fields in this compound. Taking the value of the hyperfine splitting and assuming it to be proportional to the static Ho$^{3+}$ 4f magnetic moment with $2.63\,\mu$eV/$\mu_{\rm B}$ \cite{Chatterji2013nuclear}, one obtains an estimate for the static 4f moment of Ho$^{3+}$ in HoInCu$_4$ of $5.9\,\mu_{\rm B}$.

\begin{figure}
\includegraphics[width=\columnwidth]{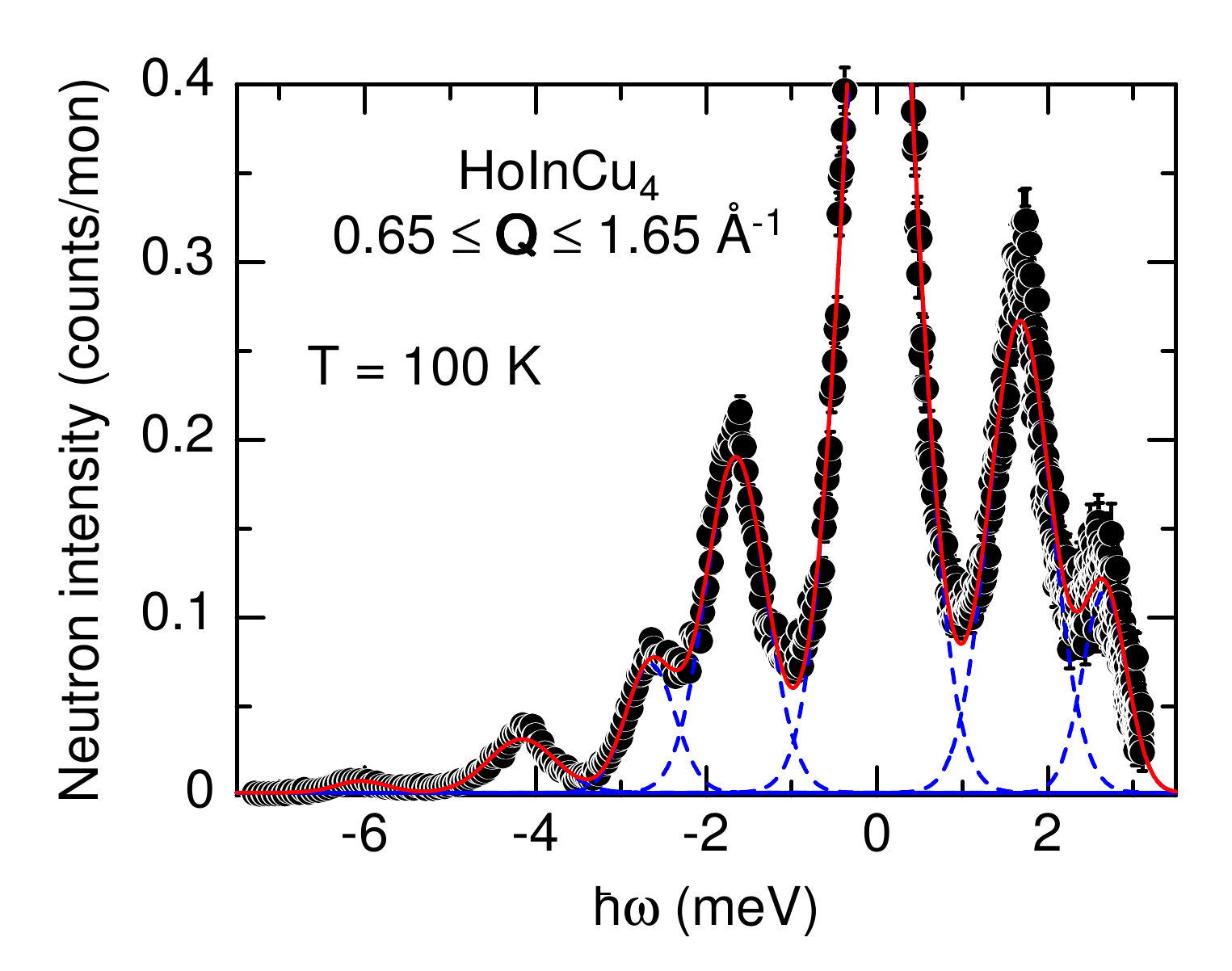}
\caption{Magnetic excitation spectrum of HoInCu$_4$ at $T = 100$\,K measured on IN6 with $E_i = 4.77$\,meV. The solid red line is the sum of fits of the inelastic peaks with gaussian lineshapes (dashed blue lines).}
\label{INS}
\end{figure}

We now turn to the determination of the CEF level scheme in HoInCu$_4$.
Figure\,\ref{INS} displays the powder INS spectrum of HoInCu$_4$ taken at $T = 100$\,K after averaging the data at low momentum transfer $0.65 < Q < 1.65$\,{\AA}$^{-1}$. Due to the large Ho$^{3+}$ moment the magnetic scattering is much stronger than any phonon scattering at this low $Q$. Therefore the spectrum shown in Fig.\,\ref{INS} can be regarded as the magnetic response. Several inelastic excitations can be clearly distinguished and have been fitted by peaks with gaussian lineshape (solid lines in Fig.\,\ref{INS}), From their momentum and temperature dependence these peaks can be attributed to CEF excitations, the most prominent peak occurring at $\hbar\omega \approx \pm1.6$\,meV. Further CEF excitations are identified at $\hbar\omega \approx 2.7$, $4.2$, and $6.0$\,meV. A Schottky anomaly in the heat capacity with a maximum in $C$ just below $10$\,K as seen in Fig.\,\ref{FullHC}\,(b) agrees well to a lowest excited CEF state at $\approx 1.6$\,meV.

In general, the crystal-field Hamiltonian can be written according to the Stevens operator formalism as
\begin{equation*}
H_{CEF} = \sum_{lm} B^m_l O^m_l({\bf J}),
\label{hcef}
\end{equation*}
where the $O^m_l({\bf J})$ denote the Stevens operators and the $B^m_l$ are the crystal field parameters. The latter include the Stevens parameters and the angular part of the wave function if we follow Hutchings convention \cite{Hutchings1964}.
For cubic symmetry the crystal field parameters are related to each other and $H_{CEF}$ simplifies to
\begin{equation*}
H_{CEF} = B_4(O^0_4 + 5O^4_4) + B_6(O^0_6 - 21O^4_6),
\label{hcef2}
\end{equation*}
According to Lea, Leask and Wolf \cite{Lea1962raising}, the two crystal field parameters $B_4$ and $B_6$ can be transformed into the two parameters $x$ and $W$ via
\begin{eqnarray*}
B_4 F(4) &=& W x \text{ and}\\
B_6 F(6) &=& W (1 - |x]),
\end{eqnarray*}
where $W$ is an energy scale factor, the parameter $x$ is limited to $|x| \leq 1$ while $F(4) = 60$ and $F(6) = 13860$ are constants.

\begin{figure}
\includegraphics[width=\columnwidth]{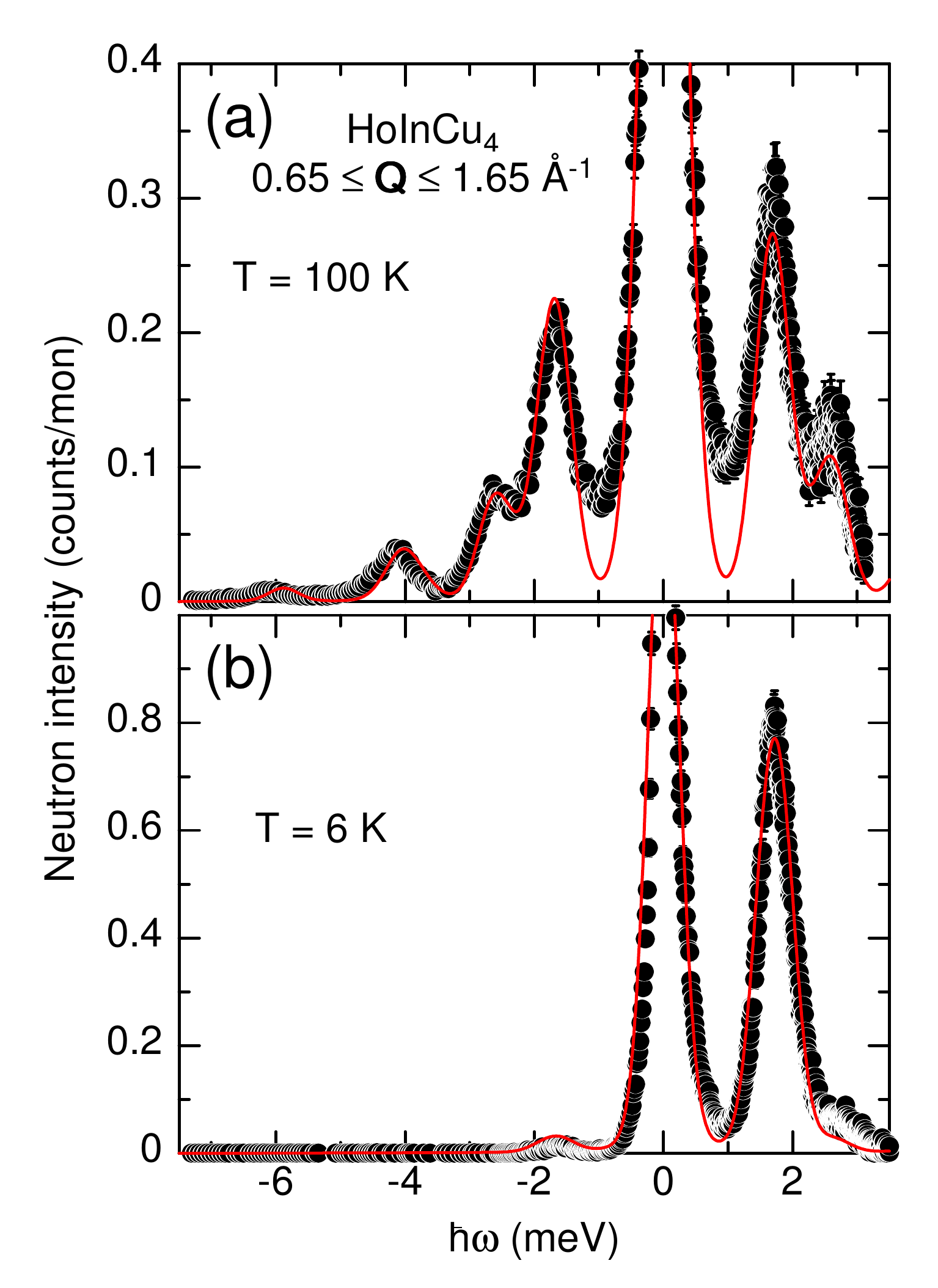}
\caption{(a) CEF fit to the inelastic magnetic excitation spectrum in HoInCu$_4$ taken at $T = 100$\,K with the best CEF parameters $x = -0.145$ and $W = 8.75\cdot 10^{-3}$\,meV (for details see text). (b) Validation of the CEF parameters obtained in the $100$\,K fits to hold as well for magnetic excitation spectrum at $T = 6$\,K.}
\label{INSfit}
\end{figure}

The Ho atoms in HoInCu$_4$ occupy the 4a site with cubic site symmetry $\overline{4}3m$. Hence, the 17 CEF states of Ho$^{3+}$ with $J = 8$ split into 4 triplets, 2 doublets and a singlet state.
 In order to describe the inelastic neutron data, all possible CEF level schemes have been calculated for the parameter range of $-1 \le x \le 1$, while $W$ has been varied only in a narrow region around the nominal value to guarantee that the first excited CEF level appears at roughly $1.6$\,meV as given by the results of INS and heat capacity mentioned above. A third fitting parameter has been introduced in modeling the $100$\,K INS data, namely a scale parameter, since the INS data were not obtained on an absolute scale. Using these three parameters the difference of calculated and measured INS spectra has been minimized (taking into account an energy broadening of the calculated CEF levels by $0.55$\,meV FWHM with a pseudo-Voigt lineshape ($90$\% gaussian shape)). The parameter range $x < -0.4$ can be ruled out from the outset, since then a nonmagnetic singlet should be the ground state \cite{Lea1962raising}. In addition, the electronic entropy gained up to $10$\,K again points to a triplet ground state. Two possible parameter regimes exist for $x$, $x \approx -0.15$ and $0.27$, yielding CEF energy levels in agreement with the observed ones. If one considers transition intensities in addition to the energies, then only the region around $x \approx -0.15$ remains as a solution for the problem. As best solution is found: $x = -0.145(15)$ and $W = 8.75(25)\cdot 10^{-3}$\,meV (the value in brackets denote the errors and indicate the $x$, $W$ values for which the sum of the squared differences of model and experimental excitation spectrum increases by 10\%) with a $\Gamma_5$ triplet ground state \cite{Lea1962raising}. This yields the following CEF level scheme: $0 (\Gamma_5) - 1.58\,\text{meV} (\Gamma_3) - 1.85\,\text{meV} (\Gamma_4) - 2.15\,\text{meV} (\Gamma_1) - 4.26\,\text{meV} (\Gamma_5) - 5.88\,\text{meV} (\Gamma_4) - 5.91\,\text{meV} (\Gamma_3)$.The $\Gamma_i$ in brackets denote the corresponding irreducible representations \cite{Lea1962raising}.
Fig.\,\ref{INSfit}(a) shows the quality of the CEF fit performed on the spectrum taken at $T = 100$\,K and indicates the validity of the fit parameters since without any further fitting the spectra taken at low temperatures can also quantitatively be well described as seen for the $6$\,K-data in Fig.\,\ref{INSfit}(b). For the CEF ground state moment a value of $4.58\,\mu_{\rm B}$ along each principal axis is expected.

\begin{figure}
\centering
\includegraphics[width=\columnwidth]{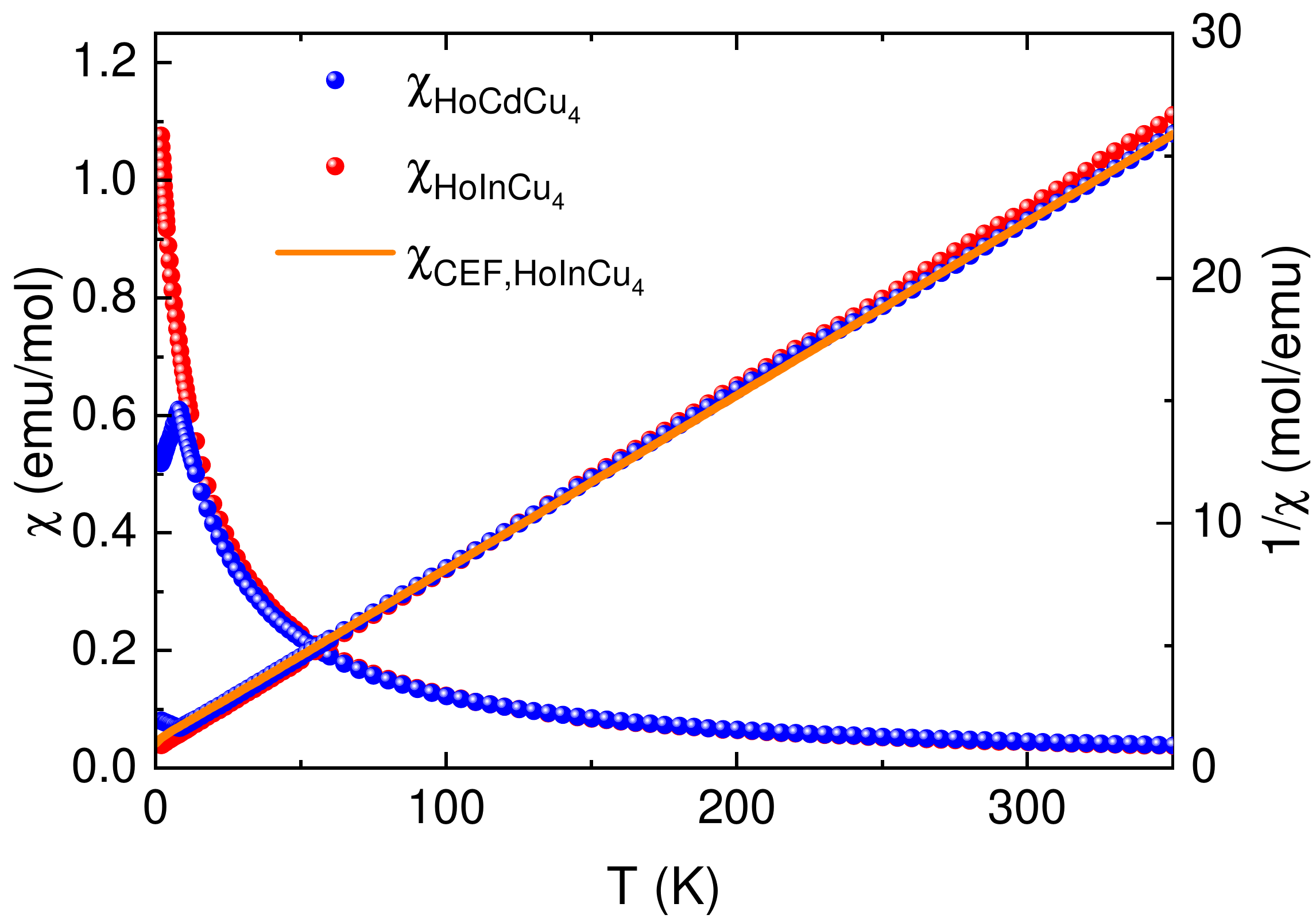}
\caption{Calculated inverse CEF magnetic susceptibility $\chi_{CEF}$ of HoInCu$_4$ (solid orange line) versus temperature. For comparison the magnetic susceptibilities of HoInCu$_4$ and HoCdCu$_4$ \cite{Fritsch2005Spin,Fritsch2006Correlation} as obtained from magnetization measurements are shown.}\label{chi}
\end{figure}

To further check the validity of the CEF parameters, the magnetic susceptibility and the heat capacity have been calculated from the CEF Hamiltonian. The solid line in Fig.\,\ref{chi} displays the calculated temperature dependence of the inverse magnetic susceptibility $\chi_{\rm CEF}^{-1}(T)$ taking into account a quite small molecular field constant of $\lambda = 1$\,mol/emu. For comparison, the experimental magnetic susceptibilities of HoInCu$_4$ and HoCdCu$_4$ reported previously \cite{Fritsch2005Spin,Fritsch2006Correlation} are also shown in Fig.\,\ref{chi}.
Calculated and experimental inverse susceptibilities coincide quite well confirming the CEF Hamiltonian. The small value of $\lambda$ rules out any sizable Kondo effect being present in HoInCu$_4$. Analyzing the high-temperature susceptibility yields for
both compounds, HoInCu$_4$ and HoCdCu$_4$, an effective paramagnetic moment $\mu_\textit{eff} = 10.5\,\mu_\textrm{B}$, which is close to the value expected from Hund's rules of $10.6\,\mu_\textrm{B}$ for Ho$^{3+}$. For HoInCu$_4$ a Weiss temperature $\theta_\mathrm{CW} = -10.9\,\si{K}$ and a N\'{e}el temperature $T_\mathrm{N} = 0.76\,\si{K}$ was found \cite{Fritsch2005Spin} (in line with heat capacity results, cf. Fig.~\ref{FullHC}\,a)), resulting in a frustration parameter $f = 14.3$, indicating a high degree of magnetic frustration in this compound \cite{Fritsch2005Spin}. The corresponding values for HoCdCu$_4$ are $\theta_\mathrm{CW} = -13\,\si{K}$, $T_\mathrm{N} = -7\,\si{K}$ and $f = 1.86$, showing that the latter compound is not frustrated at all \cite{Fritsch2006Correlation}.

\begin{figure}
\includegraphics[width=\columnwidth]{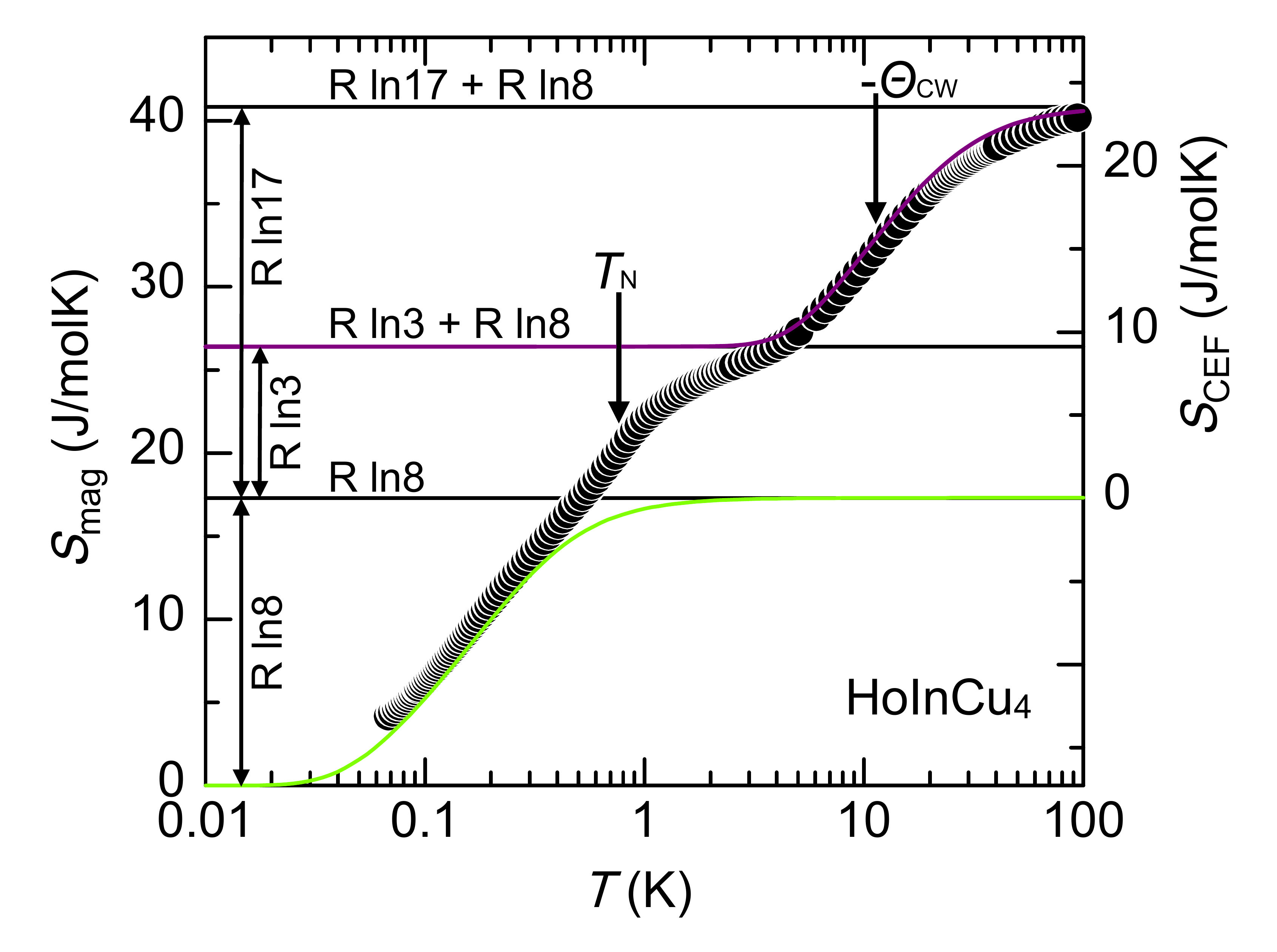}
\caption{Magnetic entropy versus temperature in HoInCu$_4$ together with the theoretical curves (same colors as in Fig.\,\ref{FullHC}). $T_N$ and $-\Theta_\mathrm{CW}$ are marked by arrows.}
\label{magnentropy}
\end{figure}

Taking a closer look at the heat capacity in HoInCu$_4$, the purple solid line in Fig.~\ref{FullHC}(b) represents the calculated contribution of the CEF excitations to the specific heat, which again agrees well with the measured data. The anomaly in $C_{mag}$ at the N\'eel temperature is quite small and rounded. In addition, the long tail in the heat capacity above $T_{\rm N}$ up to a few Kelvin evidences the existence of considerable short-range correlations of the Ho moments.
Within mean-field theory one expects a jump in the heat capacity at $T_{\rm N}$ with $\Delta C_{mag} = 2 R \approx 16.6$\,J/molK for an effective spin-1 system (the CEF ground state in HoInCu$_4$ is a triplet) \cite{Tari2003}. In contrast, an idealized jump of the measured heat capacity at $T_{\rm N}$, when neglecting the contribution of the short-range correlations, amounts to $7 - 8$\,J/molK, i.e. only about half of the expected value. This is    strong experimental evidence  that only half of the Ho moments participate in the long-range order while half of the Ho moments remain disordered (see also below). It should be mentioned that the heat capacity in HoCdCu$_4$ looks almost mean-field like at $T_{\rm N}$ (cf. Fig.\,\ref{FullHC}(c)). The discontinuity in the heat capacity at $T_{\rm N}$ is estimated to $\approx 11 - 13$\,J/mol\,K close to the theoretical expectations for HoCdCu$_4$ with all Ho moments forming long range order.

The entropy of HoInCu$_4$, determined from the heat capacity data after subtracting the phonon contribution, is shown in Fig.~\ref{magnentropy}. The entropy of the nuclear Schottky anomaly describes well the temperature dependence of the experimental data in the region of overlap and amounts roughly to $R \ln 8$, as expected. Here it should be noted that the heat capacity measurements were limited to temperatures above $T \approx 70$\,mK. Hence, experimentally less than $R \ln 8$ is gained at low temperatures and the entropy offset of the data has been adjusted by taking the extrapolation to $T = 0$ using the result of the nuclear Schottky fit. The magnetic entropy associated with the $\Gamma_5$ CEF ground state triplet is fully recovered only at temperatures well above $T_\mathrm{N}$. Right at $T_\mathrm{N}$ only one third of the expected value of $R \ln 3$ is found. This is a further proof of the partial frustration in HoInCu$_4$.  As demonstrated by the purple line in fig.~\ref{magnentropy} the data at high temperatures are very well described by the CEF calculations. Thus HoInCu$_4$ is a candidate for a classical spin liquid in the temperature range above $T_\mathrm{N}$.

\section{Frustrated magnetic Structure}
\begin{figure}
\includegraphics[width=\columnwidth]{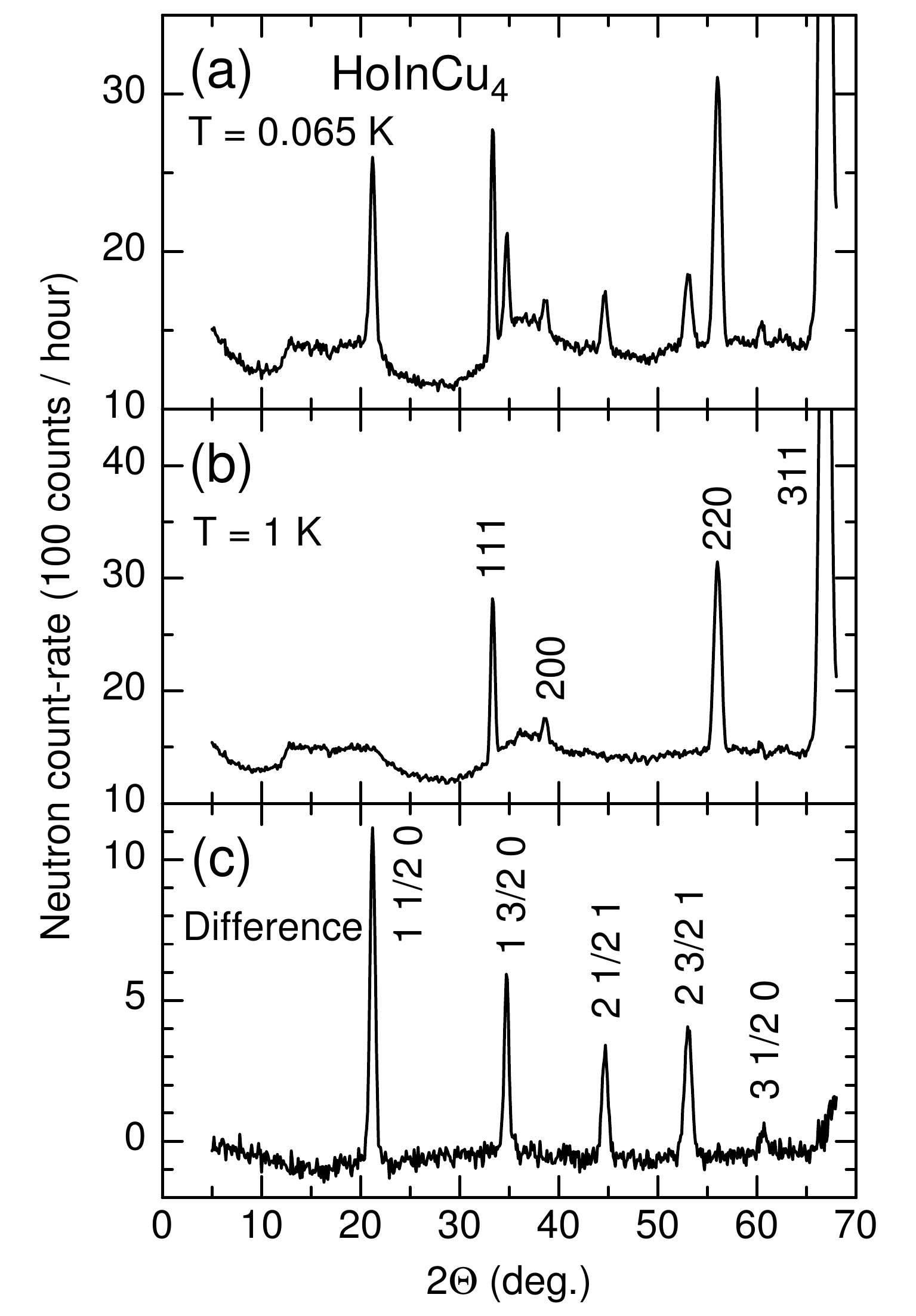}
\caption{Neutron powder diffraction pattern of HoInCu$_4$ taken at (a) $0.065$ and (b) $1$\,K. (c) Difference pattern $I(T = 0.065$\,K$)-I(T = 1$\,K$)$.}\label{diffractograms}
\end{figure}

\begin{figure}
\includegraphics[width=\columnwidth]{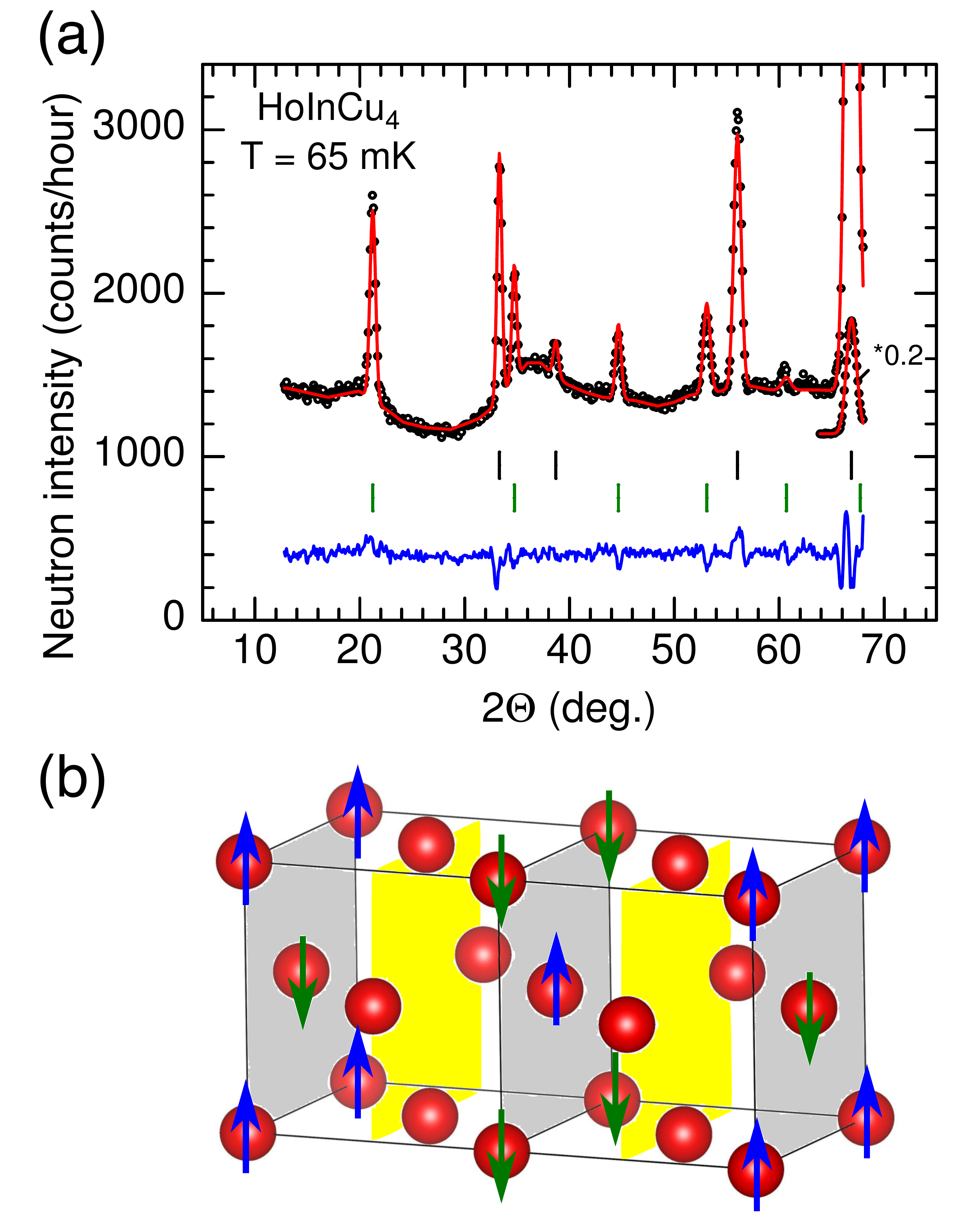}
\caption{(a) Neutron powder diffraction pattern of HoInCu$_4$ in the ground state at $T = 0.065$\,K. The red solid line indicates the fit of the crystal and magnetic structure to the data while the blue line (shifted by $400$\,counts/mon vertically) denotes the deviation of the fit from the experimental data. The two rows of ticks show the position of the nuclear Bragg peaks (top row in black) and the magnetic peaks (middle row in green) corresponding to the propagation vector ${\bf k} = \left(1\,\nicefrac{1}{2}\,0\right)$. (b)  Partially disordered magnetic structure of HoInCu$_4$ (only Ho atoms are shown).Holmium atoms located on the frustrated planes(shown in yellow), carry a magnetic moment which remains disordered below the N\'eel temperature.}\label{fpfit}
\end{figure}

Neutron powder diffraction pattern of HoInCu$_4$ taken at $T = 65$\,mK, i.e. well below $T_{\rm N}$, and in the paramagnetic state at $T = 1$\,K are displayed in Figs.\,\ref{diffractograms}(a) and (b). At $1$\,K all peaks can be attributed to the cubic crystal structure of HoInCu$_4$. Since the HoInCu$_4$ powder was immersed in deuterated methanol-ethanol mixture within the copper sample container to enhance the thermal coupling to the cold stage of the dilution cryostat, the structured background originates from diffraction of the solidified methanol-ethanol (just using helium exchange gas to thermally couple the powder to the cold state was not sufficient to cool the HoInCu$_4$ powder below $T_{\rm N}$). At lowest temperature additional peaks are clearly visible. To receive just the magnetic part of the scattering, the $1\,\si{K}$ data were subtracted from the low temperature data and are shown in Fig.\,\ref{diffractograms}(c). All magnetic peaks can be unambiguously indexed with a propagation vector $k = \left(1\,\nicefrac{1}{2}\,0\right)$ indicating a type-III antiferromagnetic structure of the magnetic fcc holmium lattice. Fullprof has been used to fit the nuclear and magnetic structure of HoInCu$_4$ at $T = 65$\,mK. The result of the fit and its deviation from the experimental data are displayed in Fig.\,\ref{fpfit} together with the positions of nuclear and magnetic Bragg peaks (vertical ticks in Fig.\,\ref{fpfit}). As seen, the  nuclear structure is well described as the cubic F$\overline{4}$3m structure determined by x-ray diffraction (see above), while the magnetic scattering indicates a type-III antiferromagnetic structure.  The fit yields for the Ho 4f moments $m = 3.23(4)\,\mu_{\rm B}$ along $[001]$, a value being much lower than expected for the $\Gamma_5$ triplet ground state. However, assuming a partially disordered state with half of the Ho moments being frustrated and carrying no ordered moment, results in $m = 4.57(6)\,\mu_{\rm B}$ per ordered Ho moment agreeing well with the CEF calculations which yield $4.58\,\mu_{\rm B}$ as noted above. Fig.\,\ref{fpfit}(b) depicts the proposed magnetic structure of HoInCu$_4$. Moment-carrying planes of antiferromagnetically ordered Ho atoms (shown in grey) are separated by planes of disordered, frustrated Ho atoms (in yellow). In fact, from symmetry considerations such a partially disordered state is allowed for the type-III antiferromagnetic structure. As already pointed out by P. W. Anderson \cite{Anderson1950} and later by J. Villain \cite{Villain1959} the four ferromagnetic sublattices of the type-III structure split into two pairs of sublattices, each pair of sublattices being antiferromagnetically ordered and carrying magnetic moments independent of the other pair. In our case just one pair of sublattices is antiferromagnetically ordered, while the other pair does not exhibit any long-range magnetic order at low temperatures.
These moments (located on the yellow planes in Fig.~\ref{fpfit}b and labeled $3$ and $4$ in Fig. 1c in ref.~\cite{Sun2018J1}) indeed carry a finite moment in our case of \mbox{HoInCu$_4$}, but they do not participate in the long-range order. The fact that all Ho atoms carry a moment can be seen e.g. from the heat capacity at very low temperatures (cf. Fig.~\ref{FullHC}b), because the nuclear Schottky anomaly indicates that all Ho atoms must have a moment. Otherwise this nuclear Schottky anomaly would be only half in size (half the $C$ value). However, the small jump height at $T_\mathrm{N}$ tells us that only half of the Ho moments order in a long-range antiferromagnetic structure. We believe that the remaining half of the Ho moments exhibit some short-range order as given by the tail in the heat capacity above $T_\mathrm{N}$.

\begin{figure}
\includegraphics[width=\columnwidth]{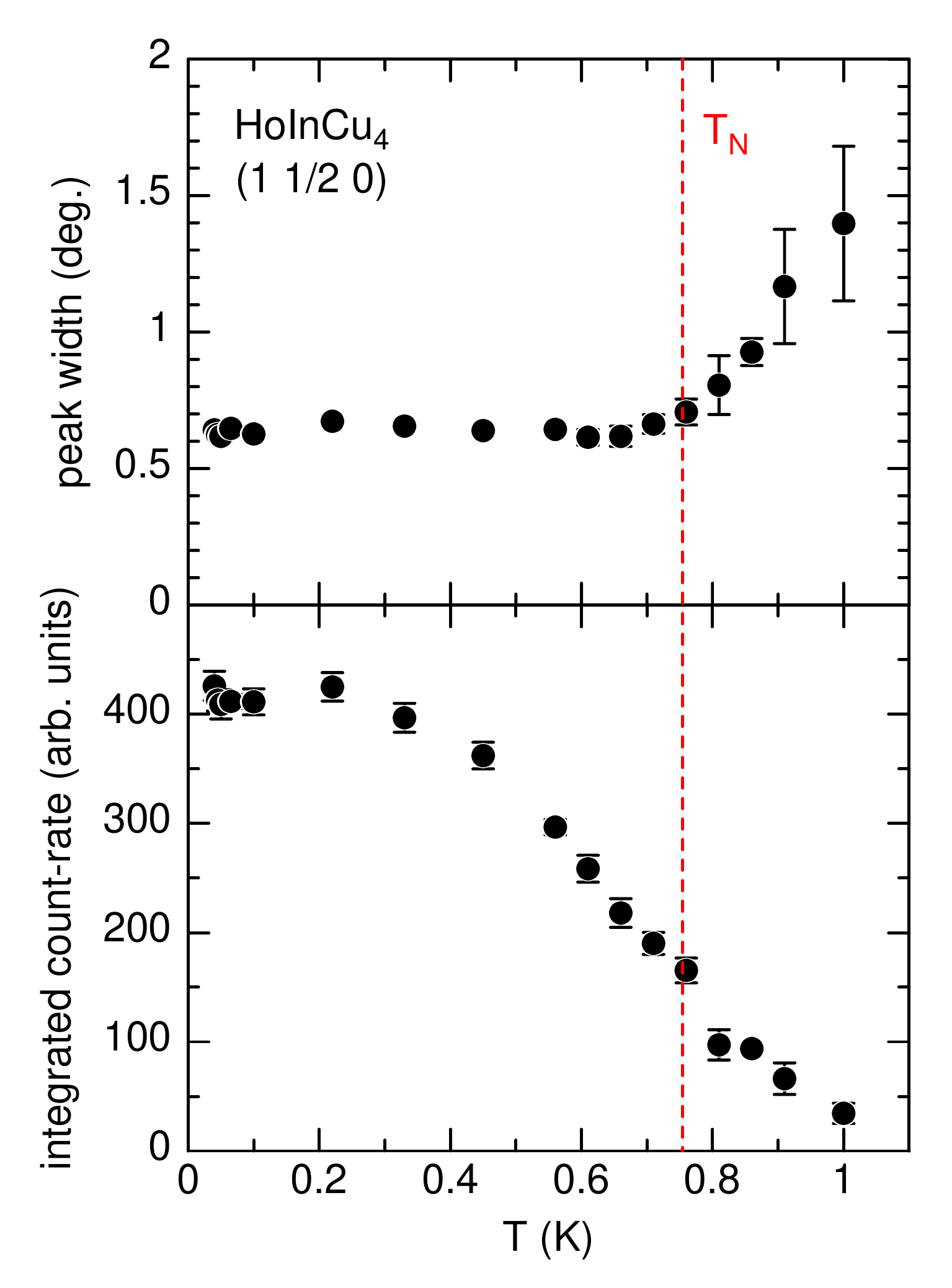}
\caption{Temperature dependence of the integrated intensity and the linewidth of the magnetic $\left(1\,\nicefrac{1}{2}\,0\right)$ superstructure peak as a result of gaussian fits to the powder diffraction pattern around $2\Theta \approx 21.2$\,deg. taken at different temperatures.}\label{intensity}
\end{figure}

\begin{figure}
\includegraphics[width=\columnwidth]{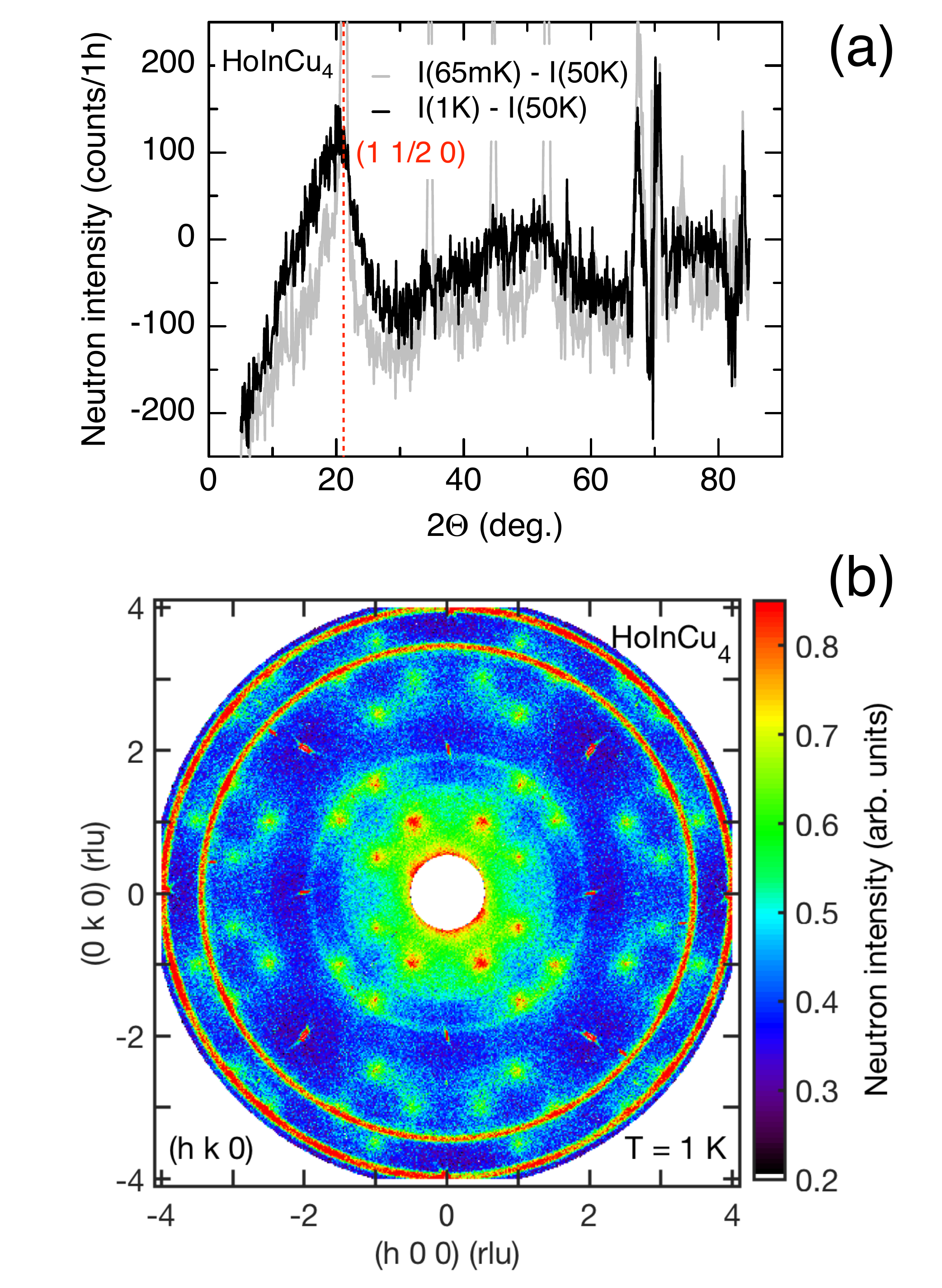}
\caption{(a) Difference neutron powder diffraction pattern of HoInCu$_4$ at $T = 65$\,mK and $1$\,K after subtracting the powder pattern recorded at $T= 50$\,K.  For the $65$\,mK pattern the intensity range is limited to small intensities to just show the diffsue signal (the magnetic Bragg peaks are truncated). (b) Intensity map of the reciprocal $(h~k~0)$ plane in single-crystalline HoInCu$_4$ taken at $T = 1$\,K. Along $l$ ($c^*$) data between $-0.1 \le l \le 0.1$\,rlu have been integrated.}\label{corr}
\end{figure}

The effect of frustration becomes also evident when looking at the temperature dependence of the magnetic order. Fig.\,\ref{intensity} displays the intensity and the linewidth of the magnetic $\left(1\,\nicefrac{1}{2}\,0\right)$ superstructure peak as function of temperature.
The magnetic intensity does not completely vanish at the N\'eel temperature, instead a considerable amount (more than $\nicefrac{1}{3}$ of the magnetic intensity for $T \rightarrow 0$) persists into the paramagnetic state. However, the peak width clearly increases above $T_{\rm N}$ indicating the break-down of long-range order.
The distinct deviations from a mean-field behavior and the considerable amount of magnetic scattering above $T_{\rm N}$ indicate the presence of strong magnetic fluctuations in the paramagnetic regime as a result of the frustration in the compound.
To further study the evolution of the magnetic correlations above $T_{\rm N}$  powder diffraction pattern were taken up to $T = 50$\,K. Fig.\,\ref{corr}(a) shows the difference powder pattern between $1$\,K and $50$\,K indicating the strong magnetic correlations in the paramagnetic state of HoInCu$_4$. Similar diffuse scattering is seen in the antiferromagnetic state as displayed in the difference pattern between $65$\,mK and $50$\,K. The diffuse intensity at $65$\,mK is direct evidence of frustration also in the antiferromagnetic state, but is weaker in comparison to the $1$\,data, since at $65$\,mK only the disordered holmium moments contribute to the diffuse intensity, while half of the holmium moments exhibit long-range order (cf. Figs.\,\ref{diffractograms} and \ref{fpfit}).
Since at $T = 50$\,K no appreciable magnetic correlations are remaining, the magnetic intensity in the $50$\,K powder pattern just follows the Ho$^{3+}$ form factor and is the reason for the negative values in the difference plot shown in Fig.\,\ref{corr}(a).  It should be mentioned that the strongest magnetic correlation peak is not centered around the $2\Theta$ position of the first magnetic superstructure peak, the $\left(1\,\nicefrac{1}{2}\,0\right)$ peak. In comparison, its position seems to be shifted to sligthly lower $2\Theta$ values (cf. Fig.\,\ref{corr}) suggesting that the spin correlations have a
different $Q$ value than the magnetic order.
A similar behavior has been reported in GdInCu$_4$ \cite{Nakamura1999partially} where the $2\Theta$ position is more compatible with simple antiferromagnetic correlations at ${\bf q} = (1~0~0)$. Such an observation suggests a competition between different possible magnetic ordering wave vectors due to competing interactions.

To answer the question if different competing interactions are present in HoInCu$_4$ and if there are indeed spin correlations with ${\bf q} = (1~0~0)$, we recorded intensity maps of the reciprocal $(h~k~0)$ plane using a HoInCu$_4$ single crystal. Fig.\,\ref{corr}(b) displays the $(h~k~0)$ intensity map taken at $T = 1$\,K, i.e. in the paramagnetic state just above $T_{\rm N}$. Apart from the two strong powder rings due to the copper sample holder and the sharp nuclear Bragg peaks, broad magnetic peaks are seen around the positions connected to the propagation vector $k = \left(1\,\nicefrac{1}{2}\,0\right)$. However, no intensity is found at $(1~0~0)$ and equivalent positions. Instead, magnetic intensity is distributed around the magnetic Bragg positions, not isotropically, but seems to connect adjacent magnetic Bragg positions. Powder averaging the single crystal data confirms the powder data of Fig.\,\ref{corr}(a).
The apparent distinct behavior in HoInCu$_4$ and GdInCu$_4$ can be resolved when reconsidering the published neutron powder diffraction pattern on GdInCu$_4$ \cite{Nakamura1999partially}. There, all powder pattern were fitted using peaks with Gaussian lineshape. However, it is known theoretically \cite{Eastabrook1952,Finger1994} that peaks in powder diffraction pattern become asymmetric in $2\Theta$, especially at very low $2\Theta$ values below $10^\circ$, due to axial divergence given by finite sample and detector sizes. Although such an asymmetric peak shape has clearly been measured in other experiments \cite{Pique2007,Granovsky2010,Granovsky2018} on the same diffractometer (D4 at ILL/Grenoble) used for the GdInCu$_4$ study, all data are still analyzed assuming a symmetric peak shape function \cite{Pique2007,Granovsky2010,Granovsky2018}. Assuming an asymmetric peak shape at small $2\Theta$ in the GdInCu$_4$ powder pattern data \cite{Nakamura1999partially} as expected theoretically \cite{Eastabrook1952,Finger1994} would allow to fit the magnetic intensity at $2\Theta = 2 - 5^\circ$ just using a single magnetic peak. This strongly questions the observation of diffuse magnetic intensity at $2$\,K centered at a different position than the long-range ordered peak.
Although we have in HoInCu$_4$ no evidence for a competition between different magnetic propagation vectors, our powder and single-crystal diffraction indicate the magnetic interactions to be anisotropic with strong magnetic correlations being present above $T_{\rm N}$. Such enhanced short-range spin correlations and a transition into the long-range magnetically ordered state only at quite low $T$ are expected in systems with strong frustration.

\begin{figure}
\centering
\includegraphics[width=0.9\columnwidth]{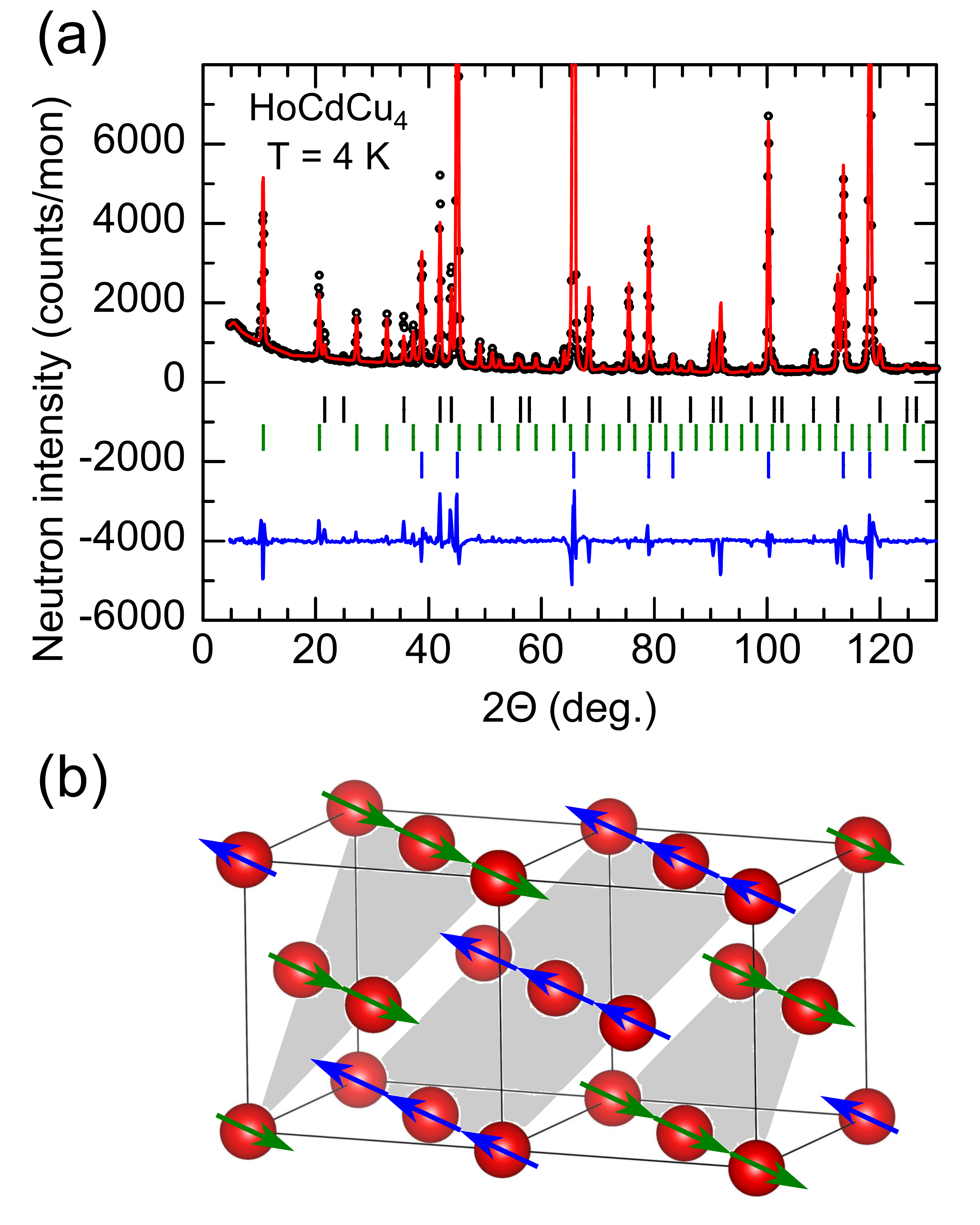}
\caption{(a) Neutron powder diffraction pattern of HoCdCu$_4$ in the antiferromagnetically ordered state at $T = 4$\,K. The red solid line indicates the fit of the crystal and magnetic structure to the data while the blue line (shifted by $4000$\,counts/mon vertically) denotes the deviation of the fit from the experimental data. The three rows of ticks show the position of the nuclear Bragg peaks (top row in black) and the magnetic peaks (middle row in green) of HoCdCu$_4$ corresponding to the propagation vector ${\bf k} = \left(\nicefrac{1}{2}\,\nicefrac{1}{2}\,\nicefrac{1}{2}\right)$. The bottom row of blue ticks corresponds to the nuclear peaks of the aluminium sample can. (b) Magnetic structure of HoCdCu$_4$ (only Ho atoms are shown).}\label{hocdcu4}
\end{figure}

For comparison, the magnetic structure in HoCdCu$_4$ has also been determined by powder neutron diffraction. Fig.\,\ref{hocdcu4} shows the powder pattern of HoCdCu$_4$ recorded at $T = 4$\,K in the magnetically ordered state. Apart from the nuclear peaks originating from the cubic HoCdCu$_4$ crystal structure (confirming the cubic C15b structure with space group F4$\overline{3}$m) and the Bragg peaks of the aluminium sample container additional magnetic superstructure peaks are visible. The magnetic peaks are absent above $T_{\rm N}$ at $T = 20$\,K and can be indexed by a propagation vector $k = (\nicefrac{1}{2}~\nicefrac{1}{2}~\nicefrac{1}{2})$. From a symmetry analysis of the possible magnetic structures compatible with the observed propagation vector the fits yield a simple type-II antiferromagnetic structure with magnetic moments $m = 9.5(2)\,\mu_{\rm B}$/Ho along $[1\overline{1}0]$, i.e. perpendicular to the propagation vector $k = (\nicefrac{1}{2}~\nicefrac{1}{2}~\nicefrac{1}{2})$. In contrast to HoInCu$_4$ all holmium moments are magnetically ordered.
No sizable spin fluctuations have been detected in the paramagnetic state of HoCdCu$_4$ indicating the absence of frustration in this isostructural system to HoInCu$_4$.

Our neutron scattering measurements are in line with heat capacity results (cf. Figs.\,\ref{FullHC}(a)+(c)) which also do not yield any indications of frustration in HoCdCu$_4$. When analyzing the magnetic contribution to the heat capacity in HoCdCu$_4$ in a similar way as HoInCu$_4$ (see above) and plotted in Fig.\,\ref{FullHC}(c), we notice that the Schottky anomaly due to the nuclear holmium moments peaks at $T = 0.25 - 0.30$\,K, a much higher temperature than in HoInCu$_4$. For the nuclear heat capacity the same analysis as for HoInCu$_4$ yields an energy splitting $a = 0.27$\,K($\simeq 2.33\cdot 10^{-2}$\,meV) of the eigenvalues $E_m$ and $p = 1 \cdot 10^{-3}$\,K($\simeq 8.6 \cdot 10^{-5}$\,meV). The calculated nuclear heat capacity of HoCdCu$_4$ using these parameters is plotted by the green solid line in Fig.\,\ref{FullHC}(c). Further calculating the static holmium 4f moment from this energy splitting by using again the linear relation \cite{Chatterji2013nuclear} yields a value of $8.9\,\mu_{\rm B}$ corroborating the large ordered moment obtained in the fits of the neutron diffraction pattern.

It should be mentioned that the strong neutron absorption prevents us from a direct determination of the CEF excitations in HoCdCu$_4$ using inelastic neutron scattering. However, as seen in Fig.\,\ref{FullHC}(c), the Schottky anomaly in the heat capacity of HoCdCu$_4$ due to CEF excitations appears at much lower temperature than in HoInCu$_4$ already indicating a drastically changed CEF level scheme with considerably smaller energy splitting in HoCdCu$_4$. Although the CEF excitation energies cannot be determined by the heat capacity measurement due to the seven different CEF levels, assuming the CEF energies in HoCdCu$_4$ to be only half of them in HoInCu$_4$ gives a fair description of the magnetic heat capacity as shown by the solid purple line in Fig.\,\ref{FullHC}(c). The CEF analysis is further hampered by the anomaly due to the onset of antiferromagnetic order which is superposed on the CEF Schottky anomaly. Nevertheless, the very low-lying CEF levels in HoCdCu$_4$, which are in the order of magnitude of the magnetic ordering temperature, are an additional support for the large ordered magnetic moment in HoCdCu$_4$.

\begin{figure}
\centering
\includegraphics[width=0.75\columnwidth]{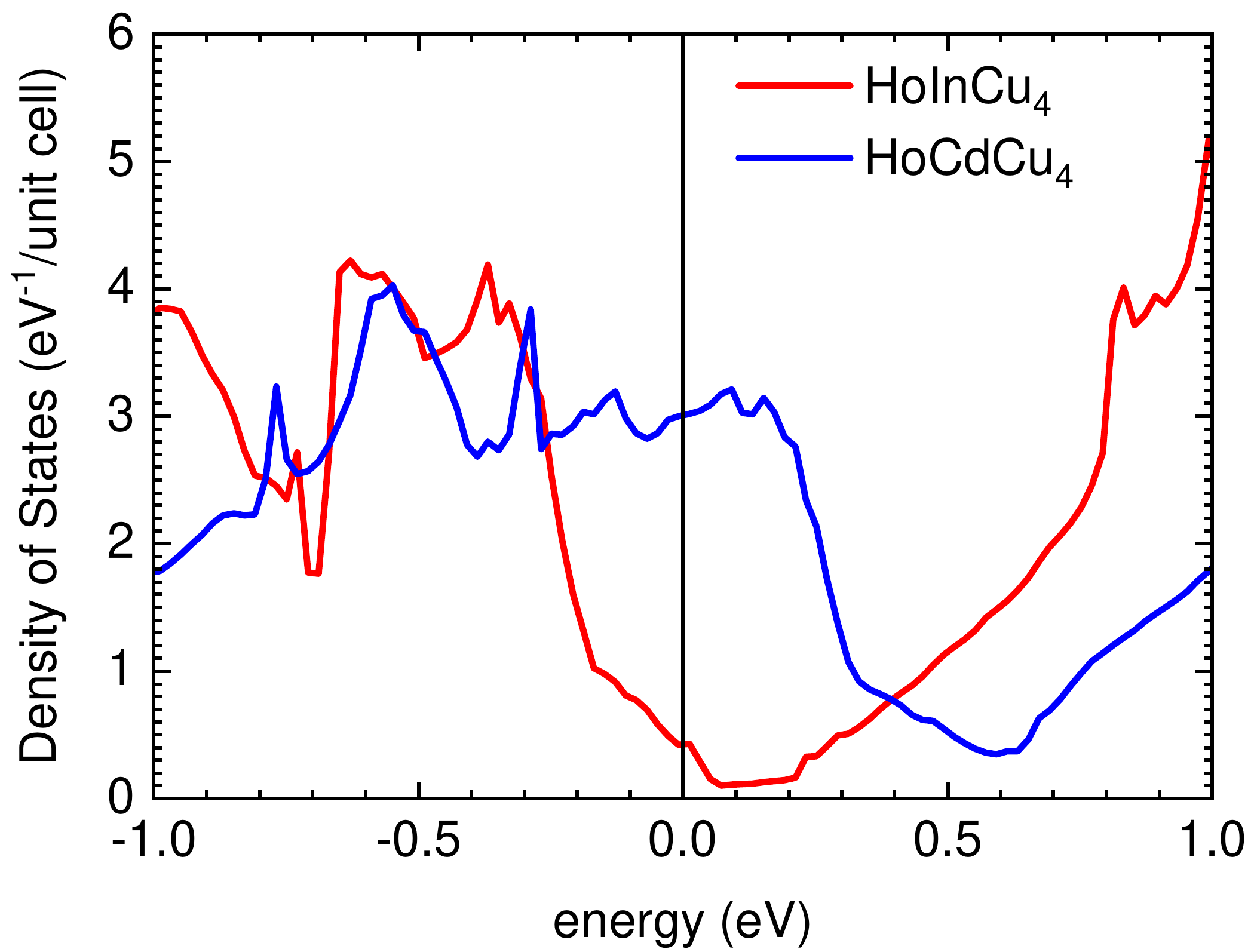}
\caption{Calculated electronic density of states of HoInCu$_4$ and HoCdCu$_4$. The holmium $4f$-electrons are considered as belonging to the core.}\label{DoS}
\end{figure}

In order to explain the difference in the magnetic structures of HoInCu$_4$ and HoCdCu$_4$, as displayed in Figs.\,\ref{fpfit}\,(b) and \ref{hocdcu4}(b), the electronic density of states was calculated. The electronic density of states (DOS) for HoInCu4 and HoCdCu4 was obtained from density-functional calculations performed in the FPLO code \cite{Kopernik1999Full} using a $k$-mesh with 1728 points in the symmetry-irreducible part of the Brillouin zone and is based on a local density approximation (LDA) for the exchange-correlation potential \cite{Perdew1992Accurate}. $4f$ states of Ho were treated as core states and thus removed from the DOS.
The results are plotted in Fig.~\ref{DoS}. For HoInCu$_4$ the Fermi level is located in a quasi-gap, but is shifted out of the gap for HoCdCu$_4$.
This corroborates the earlier assumption \cite{Fritsch2006Correlation}, that in HoInCu$_4$ the magnetic interaction is a dipole-dipole interaction, in other words only the nearest-neighbor interaction $J_1$ is relevant. For rare-earth Ising-spins the $J_1$ interaction was estimated to amount to $500$ to $800\,\si{mK}$ \cite{Bramwell2001Spin}. With increasing electronic density of states at the Fermi edge the additional RKKY interaction and therefore the next-nearest neighbor interaction $J_2$ rises, yielding a breakdown of magnetic frustration and thus a magnetically fully ordered state should occur as seen in HoCdCu$_4$. To test this conjecture, a direct determination of the magnon dispersion and hence of the nearest-neighbor and next-nearest-neighbor interactions $J_1$ and $J_2$ via inelastic neutron scattering is highly desired. A rough mean-field estimation of the ratio between $J_1$ and $J_2$ for HoInCu$_4$ and HoCdCu$_4$ is given in the Supplemental Material \cite{Supplemental} at [URL will be inserted by publisher].
The information, if HoInCu$_4$ is located close the instability at $J_2/J_1 = 0$ or $J_2/J_1 = 0.5$ in the theoretically predicted phase diagrams for an fcc lattices by Lines \cite{Lines1964Green} and Sun \textit{et al.} \cite{Sun2018J1}, could as well be obtained from the evaluation of the N\'{e}el temperatures of the doping series from HoInCu$_4$ to HoCdCu$_4$, which a planned in the near future.

\section{Conclusion}
We presented a detailed determination of the characteristic local energy scales in HoInCu$_4$ and of the magnetic structures of HoInCu$_4$ and HoCdCu$_4$. While the magnetic structure of the latter is fully ordered with a propagation vector $k = (\nicefrac{1}{2}~\nicefrac{1}{2}~\nicefrac{1}{2})$, a partially frustrated magnetic structure is realized in HoInCu$_4$. These findings are in line with the assumption of $J_2 \approx 0$ in HoInCu$_4$ within a $J_1-J_2$ model due to the low charge carrier density at the Fermi edge. The nature of the transition from the frustrated magnetic structure in HoInCu$_4$ to the non-frustrated antiferromagnetism in HoCdCu$_4$ is the topic of future research. Our measurements stress the importance of combining macroscopic and microscopic measurements to investigate the effects of frustration and should pave the way for further studies on magnetic frustration in metals, in particular in fcc lattices.

\vskip3ex

\begin{acknowledgments}
We greatly acknowledge fruitful discussions with C. Geibel, M. Brando, B. Schmidt, and K. Siemensmeyer.
We thank HZB, MLZ and ILL for the allocation of beamtime at their neutron sources and especially their technical staff for their help in making the experiments successful. This work was supported by the Freistaat Bayern through the
Programm f\"{u}r Chan\-cen\-gleich\-heit f\"{u}r Frauen in Forschung und Lehre.
Work at UD was supported by the U.S. Department of Energy, Office of Science, Basic Energy Sciences, under Award \# DE-SC0008885.
\end{acknowledgments}

\nocite{Wilson1999International}


\begin{thebibliography}{43}
\bibitem [{CCD()}]{CCDCFIZ}%
  \BibitemOpen
  \href@noop {} {}\bibinfo {note} {{The corresponding crystallographic
  information files (CIF) has been deposited with the joint CCDC/FIZ Karlsruhe
  depository with a CSD number 1914891.}}\BibitemShut {Stop}%
\end{thebibliography}

%

\end{document}


\title{Supplemental Material: Magnetic frustration in a metallic fcc lattice}

\author{Oliver Stockert}
\email{oliver.stockert@cpfs.mpg.de}
\affiliation{Max-Planck-Institut f\"ur Chemische Physik fester Stoffe, 01187 Dresden, Germany}
\author{Jens-Uwe Hoffmann}
\affiliation{Helmholtz-Zentrum Berlin f\"ur Materialien und Energie, 14109 Berlin, Germany}
\author{Martin M\"uhlbauer}
\author{Anatoliy Senyshyn}
\affiliation{Heinz Maier-Leibnitz Zentrum, 85747 Garching, Germany}
\author{Michael~M.~Koza}
\affiliation{Institut Laue-Langevin, 38042 Grenoble, France}
\author{Alexander A. Tsirlin}
\author{F. Maximilian Wolf}
\author{Sebastian Bachus}
\author{Philipp Gegenwart}
\affiliation{Experimental Physics VI, Center for Electronic Correlations and Magnetism, Institute of Physics, University of Augsburg, 86135 Augsburg, Germany}
\author{Roman Movshovich}
\affiliation{MPA-CMMS, Los Alamos National Laboratory, Los Alamos, New Mexico 87545, USA}
\author{Svilen Bobev}
\affiliation{Department of Chemistry and Biochemistry, University of Delaware, Newark, DE 19716, USA}
\author{Veronika Fritsch}
\email{veronika.fritsch@physik.uni-augsburg.de}
\affiliation{Experimental Physics VI, Center for Electronic Correlations and Magnetism, Institute of Physics, University of Augsburg, 86135 Augsburg, Germany}

\date{\today}

\maketitle
\section{Powder X-ray diffraction}
\begin{figure}[h!]
\includegraphics[width=\columnwidth]{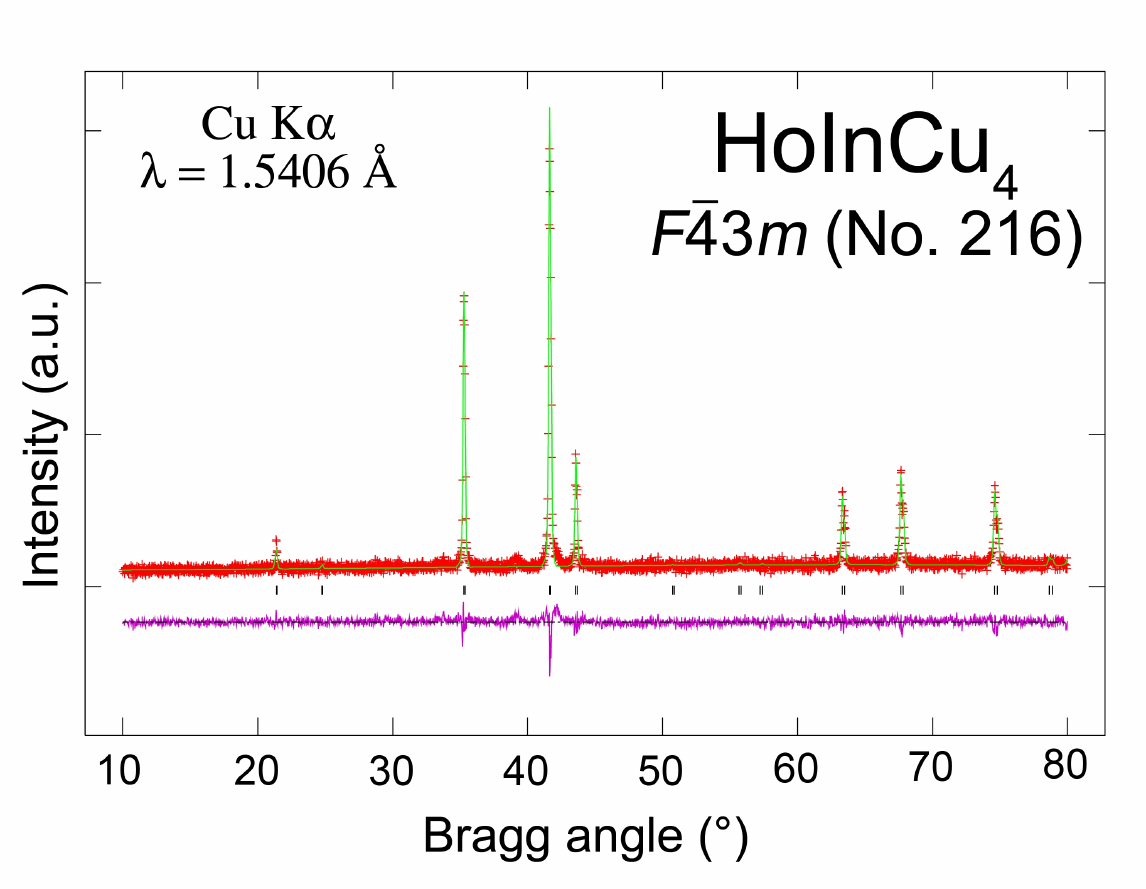}
\caption{Rietveld refinement of HoInCu$_4$: observed intensity (red), calculated (green) and difference between the two (purple).}\label{fig:xrays}
\end{figure}

\section{Single-crystal X-ray diffraction}

Intensity data sets were collected for a flux-grown crystal of HoInCu$_4$.  The crystal was cut to the desired smaller dimensions ($0.05 \times 0.04 \times 0.04\,\si{mm}$) and was then mounted on a glass fiber using Paratone N oil.  Data acquisitions took place on a Bruker SMART APEX CCD diffractometer equipped with a state-of-the-art low-temperature apparatus ($120\,\si{K}$) and using graphite monochromatized Mo K$_\alpha$ radiation.  A full sphere of reciprocal space data was collected in $3$ batch runs at different $\omega$ and $\phi$ angles with an exposure time of $10\,\si{sec/frame}$.  A total of $597$ reflections ($2 \theta_{max} \approx 56^\circ$) were collected, $66$ of which were unique ($T_{min}/T_{max} = 0.214/0.276$, $R_{int} = 0.037$).  The data collection, data reduction and integration, as well as refinement of the cell parameters were carried out using SMART and SAINT programs, respectively (SMART NT, version 5.63; Bruker Analytical X-ray Systems, Inc.: Madison, WI, 2003 / SAINT NT, version 6.45; Bruker Analytical X-ray Systems, Inc.: Madison, WI, 2003). Semi-empirical absorption correction was applied with the aid of the SADABS software package (SADABS NT, version 2.10; Bruker Analytical X-ray Systems, Inc.: Madison, WI, 2001).

The structure was subsequently solved by direct methods and refined on $F^2$ ($8$ parameters) with the aid of the SHELXTL package (SHELXTL, version 6.12; Bruker Analytical X-ray Systems, Inc.: Madison, WI, 2001).  All atoms were refined with anisotropic displacement parameters with scattering factors (neutral atoms) and absorption coefficients from the ``International Tables for Crystallography, Volume C''\cite{Wilson1999International}.  All sites were refined with anisotropic displacement parameters and full occupancies, which confirm an ordered structure, in which the Ho, Cu, and In occupy different crystallographic sites and are not disordered.  Note that Cu ($Z = 29$) and In ($Z = 49$) have sufficiently different X-ray scattering factors \cite{Wilson1999International}, which makes them clearly distinguishable. The final Fourier map is essentially flat – the highest residual density and deepest hole are less than $1 e^-/\si{\AA^3}$.

In model refinements intended to verify that there is no disorder of any kind, the site occupancy of an individual atoms was freed to vary, while the remaining ones were kept fixed – this resulted in statistically insignificant deviations from full for all – Ho in $4a$ $(\overline{4}3m)$, Cu in $16e$  $(.3m)$, and In in $4c$ $(\overline{4}3m)$, respectively.  Further details of the data collection and structure refinements for are given in Table 1, the positional and equivalent isotropic displacement parameters are listed in Table 2, respectively.  The corresponding crystallographic information files (CIF) has been deposited with the joint CCDC/FIZ Karlsruhe depository with a CSD number 1914891.

\onecolumngrid

\begin{table}
\caption{Crystal data structure refinement parameters for HoInCu$_4$.}\label{stab:xrays}
\begin{ruledtabular}
\begin{tabular}{lr}
Empirical formula       &  HoCu$_4$In \\
Formula weigth          & $533.91$ \\
Temperature             & $120\,\si{K}$ \\
Wavelength              & $0.7103\,\si{\AA}$ \\
Crystal system          & cubic \\
Space group             & $F\overline{4}3m$ \\
Unit cell dimensions    & $a = 7.1781(14)\,\si{\AA}$ \\
Volume                  & $369.85(12)\,\si{\AA^3}$ \\
$Z$                     & $4$ \\
Density (calculated)    & $9.588\,\si{g/cm^3}$ \\
Absorption coefficient  & $49.587\,\si{mm^{-1}}$ \\
$F(000)$                & $928$ \\
Crystal size            & $0.05 \times 0.04 \times 0.04\,\si{mm^3}$ \\
Theta range for data collection & $4.92^\circ$ to $28.18^\circ$ \\
Index ranges            & $-9 \leq h \leq 9$, $-8 \leq k \leq 7$, $-9 \leq l\leq 9$ \\
Reflections collected   & $597$ \\
Independent reflections & $66~[R(int) = 0.0375]$ \\
Completeness to theta $= 28.18^\circ$ & $100.0\%$ \\
Max. and min. transmission & $0.2757$ and $0.2138$ \\
Refinement method       & Full-matrix least-squares on $F^2$ \\
Data / restraints / parameters & $66$ / $0$ / $8$ \\
Goodness-of-fit on $F^2$ &  $1.133$ \\
Final $R$ indices $[I>2\sigma(I)]$ & $R_1=0.0177$, $wR_2 = 0.0372$ \\
$R$ indices (all data)  & $R_1 = 0.0184$, $wR_2 = 0.0373$ \\
Absolute structure parameter &  $0.91(5)$ \\
Extinction coefficient  & $0.0065(6)$ \\
Largest diff. peak and hole & $0.912$ and $-0614 e.\,\si{\AA^{-3}}$ \\
\end{tabular}
\end{ruledtabular}
\end{table}

\bigskip

Symmetry transformations used to generate equivalent atoms:
\begin{align*}
& \#1 x-1,-y+1/2,-z+1/2   && \#2 -x+1/2,y-1,-z+1/2      && \#3 -x+1/2,-y+1/2,z-1 \\
& \#4 x-1/2,-y+1/2,-z+1   && \#5 x-1/2,-y+1,-z+1/2     && \#6 -x+1/2,y-1/2,-z+1 \\
& \#7 -x+1,y-1/2,-z+1/2   && \#8 -x+1/2,-y+1,z-1/2    && \#9 -x+1,-y+1/2,z-1/2 \\
& \#10 x-1,y-1/2,z-1/2    && \#11 x-1/2,y-1,z-1/2      &&  \#12 x-1/2,y-1/2,z-1 \\
& \#13 x,-y+1,-z+1        && \#14 -x+1,y,-z+1          && \#15 -x+1,-y+1,z \\
& \#16 x,y-1/2,z-1/2      && \#17 x-1/2,y,z-1/2        && \#18 x-1/2,y-1/2,z \\
& \#19 x,-y+3/2,-z+3/2    && \#20 -x+3/2,y,-z+3/2      && \#21 -x+3/2,-y+3/2,z \\
& \#22 x,y+1/2,z+1/2      && \#23 x+1/2,y,z+1/2        && \#24 x+1/2,y+1/2,z \\
& \#25 x+1/2,y+1/2,z+1    && \#26 x+1/2,y+1,z+1/2      && \#27 x+1,y+1/2,z+1/2 \\
\end{align*}

\twocolumngrid

\begin{table}
\caption{Atomic coordinates ($\times 10^4$) and equivalent isotropic displacement parameters ($\si{\AA}^2 \times 10^3$) for
HoInCu$_4$. $U(eq)$ is defined as one third of the trace of the orthogonalized $U_{ij}$ tensor.}\label{stab:xrays2}
\begin{ruledtabular}
\begin{tabular}{lllll}
            & $x$   &   $y$    &    $z$  &   $U(eq)$ \\
\hline
Ho          & $0$     &  $0$    & $0$   & $8(1)$ \\
In          & $2500$ & $2500$ & $2500$ & $9(1)$ \\
Cu          & $6246(1)$ & $6246(1)$ & $6246(1)$ & $9(1)$ \\
\end{tabular}
\end{ruledtabular}
\end{table}

\begin{table}
\caption{Bond lengths $[\si{\AA}]$ for  HoInCu$_4$.}\label{stab:xrays3}
\begin{ruledtabular}
\begin{tabular}{lr}
Ho-Cu\#1 & 2.9768(7) \\
Ho-Cu\#2 &  2.9768(7) \\
Ho-Cu\#3 & 2.9768(7) \\
Ho-Cu\#4 & 2.9768(7) \\
Ho-Cu\#5 & 2.9768(7) \\
Ho-Cu\#6  &  2.9768(7)\\
Ho-Cu\#7 &   2.9768(7)\\
Ho-Cu\#8 &   2.9768(7)\\
Ho-Cu\#9  &  2.9768(7)\\
Ho-Cu\#10  &  2.9768(7)\\
Ho-Cu\#11  &  2.9768(7)\\
Ho-Cu\#12  &  2.9768(7)\\
In-Cu\#13  &  2.9749(6)\\
In-Cu\#14  &  2.9749(7)\\
In-Cu\#15  &  2.9749(6)\\
In-Cu\#4  &  2.9749(6)\\
In-Cu\#5  &  2.9749(7)\\
In-Cu\#6   & 2.9749(6)\\
In-Cu\#7   & 2.9749(7)\\
In-Cu\#8   & 2.9749(6)\\
In-Cu\#9  &  2.9749(6)\\
In-Cu\#16  &  2.9749(6)\\
In-Cu\#17  &  2.9749(6)\\
In-Cu\#18   & 2.9749(7)\\
Cu-Cu\#13   & 2.529(3)\\
Cu-Cu\#14  &  2.529(3)\\
Cu-Cu\#15   & 2.529(3)\\
Cu-Cu\#19   & 2.547(3)\\
Cu-Cu\#20   & 2.547(3)\\
Cu-Cu\#21  &  2.547(3)\\
Cu-In\#22   & 2.9750(6)\\
Cu-In\#23   & 2.9750(7)\\
Cu-In\#24   & 2.9750(6)\\
Cu-Ho\#25   & 2.9768(7)\\
Cu-Ho\#26   & 2.9768(7)\\
Cu-Ho\#27   & 2.9768(7)\\
\end{tabular}
\end{ruledtabular}
\end{table}

\FloatBarrier

\section{Mean-field estimation of exchange constants}

On the mean-field level, exchange couplings in HoInCu$_4$ and HoCdCu$_4$ can be estimated as follows \cite{Anderson1950}. The Weiss temperature of an fcc antiferromagnet is related to the sum of $J_1$ and $J_2$,
\begin{equation}
 \theta=-\frac{S(S+1)}{3}(12J_1+6J_2)=-4\tilde J_1-2\tilde J_2,
\end{equation}
where we use $\tilde J_i=S(S+1)J_i$. The N\'eel temperature $T_N$ is obtained via a similar expression,
\begin{equation}
 T_N=\frac{S(S+1)}{3}(\lambda_{12}-\lambda_{11}),
\end{equation}
where $\lambda_{11}$ and $\lambda_{12}$ stand for linear combinations of the couplings within the sublattice and between the sublattices, respectively. For the type-II order of HoCdCu$_4$,
\begin{equation}
 \lambda_{11}=6J_1,\quad \lambda_{12}=6J_1+6J_2,
\end{equation}
whereas for the type-I order of HoInCu$_4$,
\begin{equation}
 \lambda_{11}=4J_1+4J_2,\quad \lambda_{12}=8J_1+2J_2.
\end{equation}
Therefore, $T_N=\frac13(4\tilde J_1-2\tilde J_2)$ in HoInCu$_4$ and $T_N=2\tilde J_2$ in HoCdCu$_4$.

Using experimental values $\theta=-13$\,K and $T_N=7$\,K for HoCdCu$_4$, we arrive at $\tilde J_1=1.5$\,K, $\tilde J_2=3.5$\,K, and $J_2/J_1=\tilde J_2/\tilde J_1\simeq 2.33$ that should produce the type-II order indeed. On the other hand, the values of $\theta=-10.9$\,K and $T_N=0.76$\,K for HoInCu$_4$ result in $J_2/J_1\simeq 1.3$ and the type-II order again, which is now in contradiction with the experimental magnetic structure. On the other hand, mean-field theory can hardly be accurate in the strongly frustrated regime relevant to HoInCu$_4$. Assuming that short-range magnetic order appears in HoInCu$_4$ already at 3\,K (the temperature where magnetic contribution to the specific heat vanishes) and using this temperature as the mean-field $T_N$, one arrives at $\tilde J_1\simeq 2.5$\,K, $\tilde J_2\simeq 0.5$\,K, and $J_2/J_1\simeq 0.2$, which is compatible with the $\mathbf k=(1\frac120)$ order.

Whereas our mean-field estimates of $\tilde J_1$ and $\tilde J_2$ are consistent with the experimentally observed magnetic structures, the exact values should be taken with caution, given the ambiguity in the onset temperature of the short-range order in HoInCu$_4$. Moreover, we use Weiss temperatures from the high-temperature fit, but mostly the ground-state triplet contributes to the actual long-range order.